%% file: anomdim-letter.tex
\documentclass[twocolumn,preprintnumbers,superscriptaddress,aps,nofootinbib]{revtex4-2}
\usepackage[utf8]{inputenc}
\usepackage{amsmath,amssymb,mathtools}
\usepackage{graphicx}
\usepackage{booktabs,cellspace}
\usepackage{color}
\usepackage{hyperref}
\usepackage{xspace}
\usepackage[normalem]{ulem}
\usepackage{soul}
\usepackage{mathrsfs}

\makeatletter 
\renewcommand\onecolumngrid{
\do@columngrid{one}{\@ne}%
\def\set@footnotewidth{\onecolumngrid}
\def\footnoterule{\kern-6pt\hrule width 1.5in\kern6pt}%
}
\makeatother

\newcommand{\ptilde}{{\widetilde p}}
\newcommand{\qbar}{\bar q}

\newcommand{\as}{\ensuremath{\alpha_s}\xspace}

\newcommand{\np}{\text{np}}

\newcommand{\GeV}{\ensuremath{\,\mathrm{GeV}}\xspace}

\newcommand{\nc}{\ensuremath{N_\text{\textsc{c}}}\xspace}

\newcommand{\order}[1]{\mathcal{O}\!\left(#1\right)}

\newcommand{\logbook}[2]{}

\newcommand{\cR}{\mathcal{R}}
\newcommand{\cS}{\mathcal{S}}

\definecolor{darkgreen}{rgb}{0,0.4,0}
\definecolor{amaranth}{rgb}{0.9,0.17,0.31}
\definecolor{grey}{rgb}{0.5,0.5,0.5}
\definecolor{orange}{rgb}{0.9,0.5,0.0}
\definecolor{lightblue}{rgb}{0.0,0.5,1.0}
\usepackage[dvipsnames]{xcolor}

\newcommand{\CA}{C_{\!A}}

\newcommand{\OXaff}{Rudolf Peierls Centre for Theoretical Physics,
  Clarendon Laboratory, Parks Road, Oxford OX1 3PU, UK}
\newcommand{\ASCaff}{All Souls College, Oxford OX1 4AL, UK}
\newcommand{\MonashAff}{School of Physics and Astronomy, Monash
  University, Wellington Rd, Clayton VIC-3800, Australia}
\newcommand{\NewAddressCFC}{Trinity College, Dublin 2, Ireland}
\newcommand{\NewAddressRP}{St.\ Catharine’s College, Trumpington Street,\\ Cambridge CB2 1RL, UK}
\begin{document}

\title{
  Anomalous scaling of linear power corrections
}

\preprint{OUTP-25-02P}
\author{Casey Farren-Colloty}     \thanks{Current address: \NewAddressCFC}\affiliation{\OXaff}%
\author{Jack Helliwell}           \affiliation{\MonashAff}%
\author{Rtvik Patel}              \thanks{Current address: \NewAddressRP}\affiliation{\OXaff}%
\author{Gavin P.\ Salam}          \affiliation{\OXaff}\affiliation{\ASCaff}%
\author{Silvia Zanoli}            \affiliation{\OXaff}%

\begin{abstract}
  Non-perturbative corrections to hadronic observables represent a
  critical obstacle to increasing accuracy at colliders.
  Long taken to scale simply as $1/Q$, where $Q$ is the centre-of-mass
  scattering energy, recent work has opened the path towards
  calculating the anomalous dimension that modifies that scaling.
  A priori, the problem is complex, requiring a resummation involving
  arbitrary numbers of large-angle and low-energy gluons.
  Within a specific framework for kinematic recoil, we
  show that it reduces to a simple exponential for key observables
  like the thrust, $C$-parameter and energy correlators.
  This simplicity holds for a specific hadron mass scheme, and also
  even beyond the two-jet limit.
\end{abstract}
\maketitle

The interpretation of ever more precise data from high-energy
colliders requires corresponding improvements in the accuracy of
theoretical predictions within the Standard Model (SM) of particle
physics.
One major avenue of progress involves
advances~\cite{Heinrich:2020ybq,Campbell:2022qmc} in calculations
using perturbative Quantum Chromodynamics (QCD).
Yet, as perturbative accuracy increases, a second problem comes to the
fore, namely that experiments measure non-perturbative hadrons rather
than the quarks and gluons of perturbative QCD, and one must
understand the relation between the two.

One powerful and widely used approach to this problem is to use
phenomenological models to describe the transition from partons to
hadrons,
i.e.\
\emph{hadronisation}~\cite{Artru:1974hr,Bowler:1981sb,Andersson:1983jt,Andersson:1983ia,Christiansen:2015yqa,Gottschalk:1982yt,Gottschalk:1986bv,Field:1982dg,Webber:1983if}.
However, such models bring little analytical insight into the problem
and suffer from long-standing open questions about how to consistently
combine them with the highest-accuracy perturbative calculations.
An alternative approach exploits the renormalon breakdown of
perturbation theory at high orders, a factorial growth in the
perturbative coefficients~\cite{Beneke:1998ui}, to gain insight into the
potential structure of non-perturbative effects.
This path offers the potential for greater understanding of the
analytical structure of hadronisation.\footnote{In hadron-hadron
  collisions, specifically the part not related to multi-parton
  interactions.}


The class of non-perturbative corrections that is largest numerically
is that with a $\Lambda/Q$ power scaling, where $\Lambda$ is the
non-perturbative scale of QCD and $Q$ is the centre-of-mass
energy.
Such linear power corrections affect essentially all infrared safe
observables that are based on hadron momenta.
They were first explored in the
mid-1990's~\cite{Manohar:1994kq,Dokshitzer:1995zt,Akhoury:1995sp,Nason:1995np,Dokshitzer:1995qm,Beneke:1997sr,Dokshitzer:1997ew,Dokshitzer:1997iz,Dokshitzer:1998pt,Dasgupta:1998xt,Dasgupta:1999mb,Korchemsky:1999kt}
and are among the key elements that are extensively debated in
determinations of the strong coupling constant from event shapes~\cite{Nason:2023asn,Bell:2023dqs,Benitez:2024nav,Nason:2025qbx,Aglietti:2025jdj,Benitez:2025vsp}, with
tensions of up to $3$ standard deviations~\cite{Benitez:2024nav} relative
to the world average~\cite{ParticleDataGroup:2024cfk}.

\begin{figure}
  \centering
  \includegraphics[width=\columnwidth]{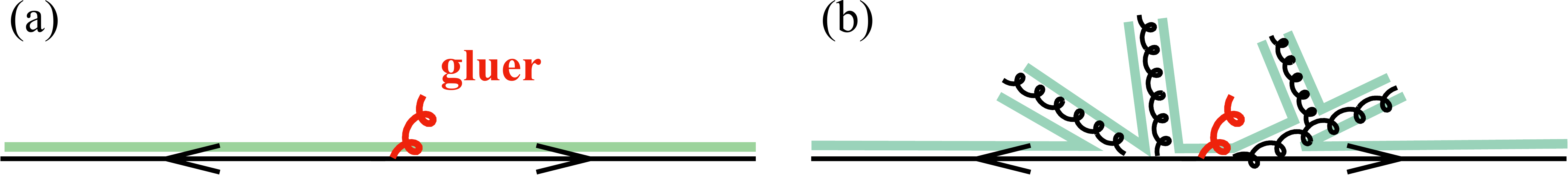}
  \caption{(a) emission of a non-perturbative \emph{gluer} from a
    $q\bar q$ system, as in standard calculations; (b) a realistic set
    of perturbative partons from which a gluer would actually be
    emitted.
  }
  \label{fig:basic-picture}
\end{figure}

One of the key results of the 1990's is that the renormalon approach
reduces to probing how the emission of a single
gluon at low transverse momentum  modifies a given observable.
That single gluon is sometimes referred to as a \emph{gluer}~\cite{Dokshitzer:1995zt}.
This is illustrated in Fig.~\ref{fig:basic-picture}a, which shows the
emission of a gluer (in red) in $e^+e^- \to q\bar q$ collisions.

Such an approach inevitably oversimplifies hadronisation, which seldom
happens from a bare $q\bar q$ system, but instead almost always from a
system where the $q\bar q$ has radiated multiple perturbative gluons,
as in Fig.~\ref{fig:basic-picture}b.
What long remained unclear, however, was how to extend the renormalon
picture to systems with more than a $q\bar q$ pair.
The past few years have started to see insight into this
question~\cite{FerrarioRavasio:2018ubr,FerrarioRavasio:2020guj,Luisoni:2020efy,
  Caola:2021kzt,
  Caola:2022vea,Makarov:2023ttq,Makarov:2023uet,Makarov:2024ijn} (for
other recent hadronisation studies, including operator approaches, see e.g.\
Refs.~\cite{Benitez:2024nav,Bell:2023dqs,Banfi:2023mes,Hoang:2024nqi,Chen:2024nyc,Benitez:2025vsp}).

In the picture that is emerging --- but that remains to be fully
established --- one can understand the effect of
hadronisation by evaluating how an observable changes when inserting a
gluer into each and every colour dipole of the event, using a
well-chosen recoil scheme to account for momentum conservation.

Developing that picture for the thrust~\cite{Brandt:1964sa,Farhi:1977sg} and
C-parameter~\cite{Ellis:1980wv}, Dasgupta and Hounat~\cite{Dasgupta:2024znl} have
considered the average effect of gluer insertion, after integrating over the
kinematics of a single perturbative soft gluon.
They find that this leads to a modification of the coefficient of
$\Lambda/Q$ for the average value of those observables, observing a
correction factor that involves an anomalous dimension,
$1 - (\as/2\pi)\, \CA\, \cS_1 \ln Q/\Lambda$, with $\cS_1 \simeq 2.455$
common to both observables.
As discussed there, the all-order resummation of the anomalous
scaling would involve considering gluer emission from systems with
arbitrary numbers of large-angle soft gluons,
Fig.~\ref{fig:basic-picture}b.
The goal of this letter is to carry out that resummation.
A priori, one expects~\cite{Dasgupta:2024znl} to find a structure
similar to that of non-global logarithms~\cite{Dasgupta:2001sh}, for
which no analytic resummation result is known.

Starting from a $q\bar q$ system, the non-perturbative correction to
an observable $V$ can be written as
\begin{multline}
  \label{eq:deltaVqq}
  \langle \delta V_\np\rangle_{q\bar q} =
  \int_0^{\mu_\np} \frac{dk_{tn}}{k_{tn}} \int d\eta_n \frac{2C_F
    \as^{(\text{eff})}(k_{tn})}{\pi}
  \times
  \\
  \times \left[V(p_1,p_2,k_n) - V(\tilde p_1,\tilde p_2)\right],
\end{multline}
where the $p_i$ ($\ptilde_i$) are the hard momenta in the event after
(before) emission of a soft non-perturbative gluer $k_n$.
The integral is performed over its transverse momentum $k_{tn}$ and
rapidity $\eta_n$, weighted with $\as^{(\text{eff})}(k_{tn})$, an
effective non-perturbative coupling for emitting the gluer at scales
below an infrared matching scale $\mu_\np\sim 1\GeV$.

We consider observables $V$ that are \emph{linear}, i.e.\ reduce to linear
functions of any number of soft massless momenta $k_i$ emitted from
the hard $q\bar q$ system
\begin{equation}
  \label{eq:V-linearity}
  V(p_1,p_2,k_1\ldots,k_m) - V(\ptilde_1, \ptilde_2) =
  \!\sum_i \frac{k_{ti}}{Q} f_V(\eta_i) + \order{\!\frac{k_{ti}^2}{Q^2}\!}\!,
\end{equation}
specifically,
the thrust ($T$, $f_\tau(\eta)=e^{-|\eta|}$, with
$\tau\equiv1-T$),
the $C$-parameter ($f_C(\eta) = 3/\cosh\eta$),
and energy-energy correlators
$\text{EEC}(\theta)=\sum_{ij} E_i E_j/Q^2
\delta(\theta_{ij}-\theta)$~\cite{Basham:1978bw,Basham:1977iq}
($f_\text{EEC}(\eta) =
2\cosh^2\!\eta\,\delta(\eta-\log(\tan\theta/2))$).
Here $k_{ti}$ and $\eta_i$ are the transverse momentum and rapidity of
$i$ with respect to the $q\bar q$ pair.
Eq.~(\ref{eq:deltaVqq}) then  reduces to
\begin{equation}
  \label{eq:2}
  \langle \delta V_\np\rangle_{q\bar q} = c_V\frac{T_{q\bar q}}{Q},
  \;\;
  T_{q\bar q} = \int_0^{\mu_\np} \!\!\!dk_{tn} \frac{2C_F
    \as^{(\text{eff})}(k_{tn})}{\pi},
\end{equation}
where $T_{q\bar q}$ can be interpreted as the effective non-perturbative transverse
momentum produced per unit rapidity from a $q\bar q$ system, and
$c_V = \int d\eta f_V(\eta)$, e.g.\ $c_\tau=2$, $c_C=3\pi$ and
$c_{\text{EEC}(|\cos\theta<1/2|)}=4/\sqrt{3}$.
The linearity of the observables ensures that the details of how that
transverse momentum is produced, e.g.\ correlations between particles,
do not matter.

The key question we ask here is how the transverse momentum per unit
rapidity changes when the non-perturbative gluer emission takes place
not from a $q\bar q$ pair, but from a system with arbitrary numbers of
additional soft gluons, $qg_1\dots g_m \bar q$.
We will work in the large-$\nc$ limit, considering
gluer emission from
a sequence of colour dipoles $qg_1$, $g_1g_2$, etc.
The essential observation of Refs.~\cite{Caola:2021kzt,Caola:2022vea}
is that for linear observables, 
the result of the renormalon calculation can be obtained using
kinematic maps where the recoil of the hard partons is a linear
function of the soft emitted gluer's momentum.
It is important to be aware that this is yet to be proved for
final-states with gluons, though the universality of soft emission
suggests that the results should carry over.

We will consider two maps that satisfy the soft-gluer linearity condition: the
$3$-jet matched~\cite{Hamilton:2023dwb}
PanGlobal~\cite{Dasgupta:2020fwr,FerrarioRavasio:2023kyg} and
PanLocal~\cite{Dasgupta:2020fwr} maps.
Both have the property that longitudinal recoil is local to the dipole.
We have verified that our implementation reproduces the $3$-jet non-perturbative
shifts of Refs.~\cite{Caola:2021kzt,Caola:2022vea,Nason:2023asn}
(supplemental material~\cite{supplement},
\S\ref{sec:supp:lit-comparisons}) and that the maps' predicted
non-perturbative shifts for linear observables agree also for more
complex dipole configurations, for sufficiently soft gluers.

\begin{figure}
  \centering
  \includegraphics[width=0.8\columnwidth]{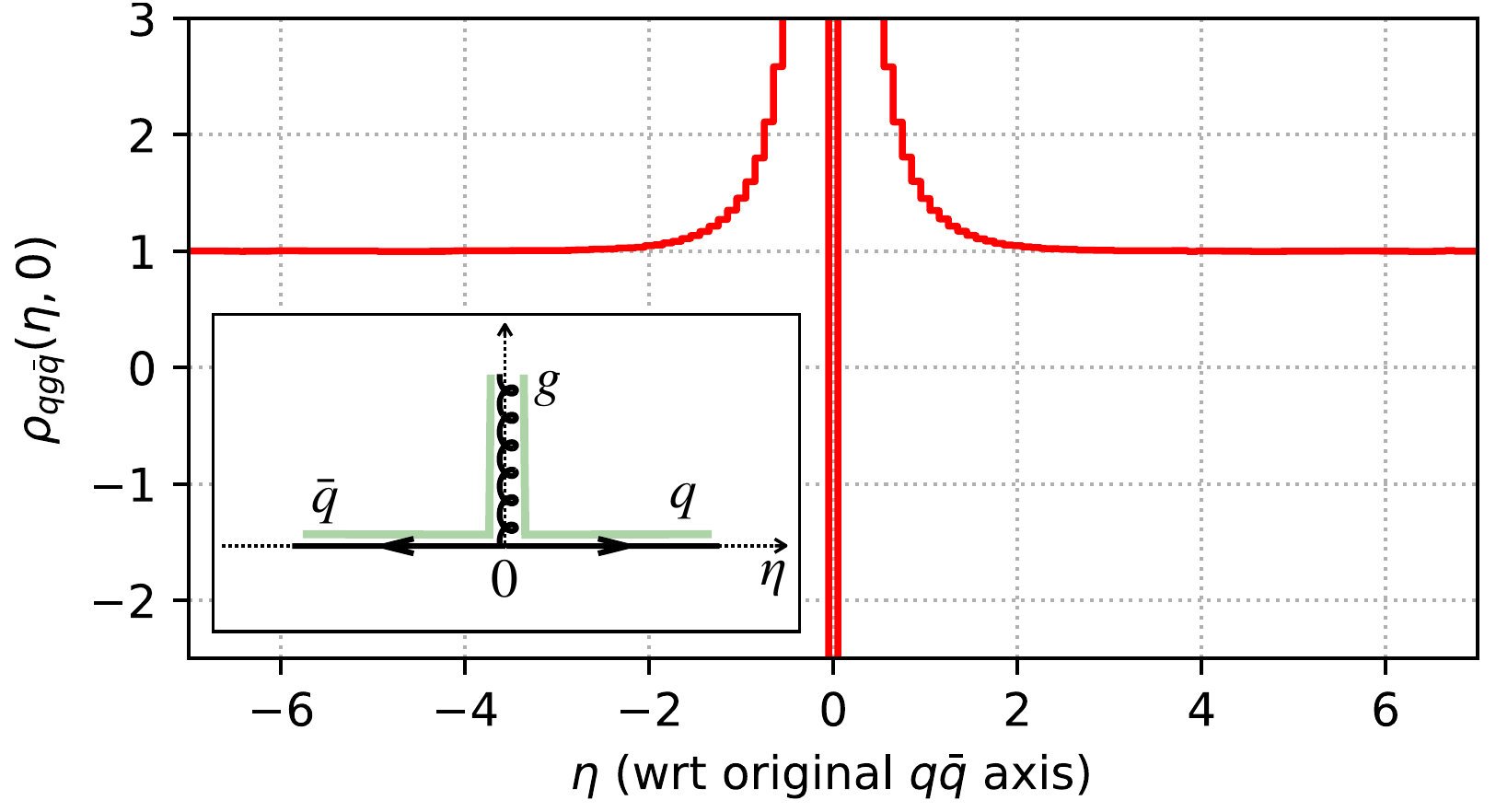 }        
  \caption{The effect of gluer emission on transverse momentum per
    unit rapidity ($\eta$) in events consisting of a hard $q\bar q$
    pair and a perturbative soft gluon at zero rapidity.  }
  \label{fig:one-emission}
\end{figure}

We start by considering a gluer emission from a $qg\bar q$ system, where
the gluon itself is soft, and
evaluate the non-perturbative transverse momentum per unit rapidity as
a function of rapidity $\eta$ relative to the $q\bar q$ system.
That can be written as 
\begin{multline}
  \label{eq:Tqqgy}
  T_{q g \bar q}(\eta,\eta_g)
  = \frac{\CA}{\pi} \sum_{qg,g\qbar}\int \frac{dz_+}{z_+} \frac{dk_{\perp n}}{k_{\perp
      n}} \frac{d\phi_n}{2\pi}
  \left[ k_{tn}\delta(\eta-\eta_n) + 
    \right.\\\left.
    + \left(p_{tg} - \ptilde_{tg}\right)\delta(\eta-\eta_g) \right]
  \as^{(\text{eff})}(k_{\perp n})\,.
\end{multline}
The square brackets contain the change in scalar transverse momentum
(subscript $t$, with respect to the $q\bar q$ direction) due to the
gluer emission $n$ and recoil of the perturbative gluon $g$.
The gluer integration variables $k_{\perp n}$ and $z_+$ are the
transverse component and plus light-cone fraction defined with respect to
the emitting $qg$ or $g\qbar$ dipole.
We neglect the transverse momenta of the $q,\bar q$, because they are
at large rapidities, where $f_V(\eta)$ in Eq.~(\ref{eq:V-linearity})
vanishes.
Finally, the gluer must be much softer than the gluon.
Rather than using some specific form for $\as^{(\text{eff})}(\mu)$, we
probe how the observable changes on insertion of the gluer at a
specific transverse momentum $k_{\perp,n}$, then taking the limit of
the gluer $k_{\perp,n} \to 0$~\cite{supplement},
\S\ref{sec:supp:lit-comparisons}. 
Fig.~\ref{fig:one-emission} shows a numerical evaluation of
\begin{equation}
  \label{eq:rho-def}
  \rho_{q g \bar q}(\eta, \eta_g) \equiv
  T_{q g \bar q}(\eta,\eta_g) \,/\, T_{q\bar q}
\end{equation}
for $\eta_g = 0$.
At large rapidities, $\rho_{q g \bar q}(\eta,\eta_g)$ tends to $1$ as expected
from angular ordering.
For $\eta$ near zero, i.e.\ close to the perturbative gluon, a small
$k_{\perp n}$ with respect to the $qg$ or $g \bar q$ dipoles
translates into a larger $k_{t n}$ with respect to the $q\bar q$
direction, hence $\rho_{q g \bar q}(\eta) \gg 1$.
Finally there is a negative $\delta$-function at $\eta=\eta_g=0$  associated
with the $\left(p_{tg} - \ptilde_{tg}\right)$ recoil of the
perturbative gluon.

To obtain the logarithmically dominant contribution to
Eq.~(\ref{eq:2}) at NLO accuracy, we need to integrate
$\rho_{q g \bar q}(\eta,\eta_g)$ over the perturbative soft gluon
kinematics and account for virtual corrections:
\begin{equation}
  \label{eq:RNLO-step}
  \cR^{\text{NLO}} = 1 + \int \!d\eta_g \!\int_{\mu_\np}^Q
  \!\frac{d\tilde p_{tg}}{\tilde p_{tg}} \frac{2C_F \as(\tilde p_{tg})}{\pi}
  [ \rho_{q g \bar q}(\eta,\eta_g) - 1].
\end{equation}
This is the NLO factor that multiplies the average transverse momentum
density.
It is straightforward to show (Ref.~\cite{supplement}, \S\ref{sec:supp:S1}) that
\begin{equation}
  \label{eq:R-integral}
  \int_{-\infty}^{+\infty} d\eta_g [ \rho_{qg\bar q}(\eta,\eta_g) - 1] = -4\left(1 - \ln 2\right)\,.
\end{equation}
Replacing $2C_F \to \CA$ (large-$\nc$ limit) yields,
\begin{equation}
  \label{eq:RNLO-final}
  \cR^{\text{NLO}} = 1 - \lambda(Q,\mu_\np) \frac{\CA \cS_1}{2\pi},
\end{equation}
with
\begin{equation}
  \label{eq:S1-and-lambda}
  \cS_1 = 8\left(1 - \ln 2\right)\,,
  \qquad
  \lambda(Q,\mu_\np) = \int_{\mu_\np}^Q \frac{dp_t}{p_t} \as(p_t)\,,
\end{equation}
in agreement with the numerical result of
Ref.~\cite{Dasgupta:2024znl}, but obtained generically for observables
that are linear in the soft momenta and with $f(\eta) \to 0$ for large
$|\eta|$. 

Our approach can be straightforwardly extended to all orders.
Specifically, let $\cR$ be a function of the
hard scale $Q$, $\cR(Q)$, such that $\cR(\mu_\np) = 1$.
Given a certain $\cR(Q)$, if we increase the hard scale by an
infinitesimal factor $1+\epsilon$, we allow for extra configurations
with a soft gluon between scales $Q$ and $(1+\epsilon)Q$. Those
extra configurations have a radiation pattern corresponding to
dressed $qg$ and $g\bar q$ dipoles, where the non-perturbative
transverse momentum density with respect to the $qg$ or
$g\bar q$ dipole is $\cR(Q) T_{q\bar q}$.
Recall that we are using the large-$\nc$ limit, thus the $qg$ and
$g\bar q$ dipoles have the same non-perturbative emission density as
a $q\bar q$ dipole.
This yields the following differential equation
\begin{equation}
  \label{eq:dRdlnQ}
  \frac{d\cR(Q)}{d\ln Q}
  = \frac{2C_F \as(Q)}{\pi} \int d\eta_g [ \rho_{qg\bar q}(\eta,\eta_g) - 1]\cR(Q)\,,
\end{equation}
where we make use of the property that the non-perturbative all-order
longitudinal recoil from the $qg$ or $g\bar q$ dipoles gets the same
$\cR(Q)$ factor as the transverse momentum density (\cite{supplement},
\S\ref{sec:supp:perp-v-long}).
Once again using the large-$\nc$ limit, $2C_F \to C_A$, the solution
to Eq.~(\ref{eq:dRdlnQ}) is
\begin{equation}
  \label{eq:Rsoln}
  \cR(Q)
  = \exp\left[- \lambda(Q,\mu_\np) \frac{\CA \cS_1}{2\pi} \right]
  = \left(\frac{\as(Q)}{\as(\mu_\np)}\right)^{\frac{\CA\cS_1}{\beta_0}}
  ,
\end{equation}
with $\beta_0 = (11 C_A - 2n_f)/3$, such that Eq.~(\ref{eq:2}) becomes
\begin{equation}
  \label{eq:deltaV-final}
  \langle \delta V_\np\rangle = c_V
  \frac{T_\text{all-order}(Q)}{Q},\quad
  T_\text{all-order}(Q)
  = 
  T_{q\bar q} \cR(Q)\,. 
\end{equation}
This is much simpler than the structure one would expect from a
non-global resummation, and part of the reason why such a structure
emerges is the requirement of linearity of the observable
in the soft limit, Eq.~(\ref{eq:V-linearity}).
The exponentiated structure is important because there is a
fundamental ambiguity in what scale one should take for $\mu_\np$, and
Eq.~(\ref{eq:Rsoln}) tells us that that ambiguity can simply be
absorbed into the normalisation of $T_{q\bar q}$.

\begin{figure}
  \centering
  \includegraphics[width=\columnwidth]{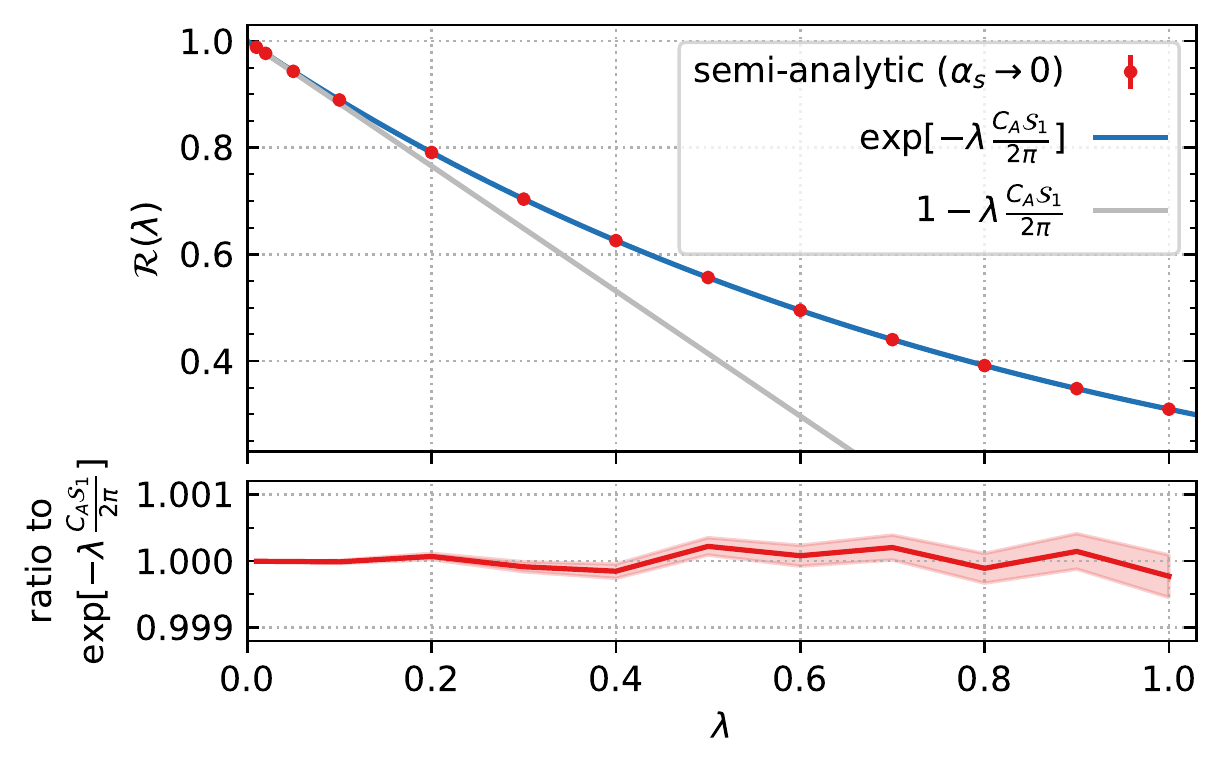}
  \caption{All-order evaluation of $\cR$, using a perturbative shower
    for the gluon cloud and a semi-analytic evaluation of the gluer insertion.
    The result is shown in the limit $\as \to 0$ at any given fixed
    $\lambda$, plotted as a function of $\lambda$ and compared to the
    all-order analytic result, Eq.~(\ref{eq:Rsoln}) (blue line).  }
  \label{fig:all-order-many-lambdaf}
\end{figure}

We supplement our analytic derivation with numerical tests.
We start with a $q\bar q$ system, and use a perturbative parton
shower~\cite{FerrarioRavasio:2023kyg,vanBeekveld:2023ivn,vanBeekveld:2025lpz} to add a
cloud of soft gluons between scales $p_{t,\min}$ and $p_{t,\max}\ll Q$
(using fixed $\as$ for simplicity).
We then have two ways of probing the gluer's effect.
In one, we directly insert the gluer into the showered event.
In the other, we use a semi-analytic calculation of the gluer's impact
on the transverse momentum per unit rapidity for an arbitrary
dipole~\cite{supplement}, \S\ref{sec:supp:semi-analytic}.
Fig.~\ref{fig:all-order-many-lambdaf} shows results with the latter.
We have run the shower in an asymptotic regime, $\as \to 0$ for
fixed $\lambda = \as \ln p_{t,\max}/p_{t,\min}$, so as to extract
just the $(\as \ln p_{t,\max}/p_{t,\min})^n$
terms, free of higher-logarithmic contamination~\cite{Dasgupta:2020fwr}.
The results agree with Eq.~(\ref{eq:Rsoln}) to within statistical errors,
of order a few times $10^{-4}$.
The comparison to the pure first order result (grey line) illustrates
the size of the resummation effect.

\begin{figure}
  \centering
  \includegraphics[width=0.93\columnwidth]{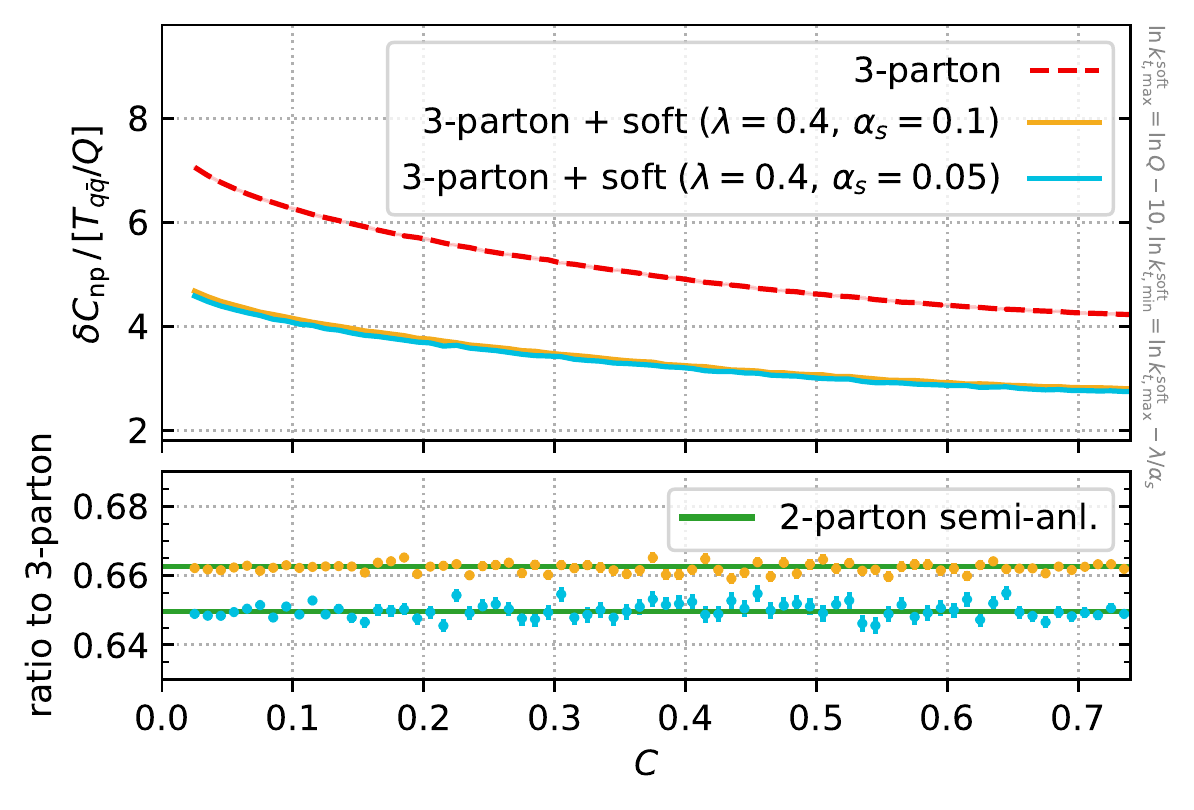}
  \caption{The $C$-parameter non-perturbative correction in the
    $3$-jet region, without (red) and with a cloud of soft gluons (orange, cyan).
    The lower panel shows that the ratio with and without soft
    gluons is independent of $C$ and agrees with the
    corresponding $2$-jet result at the same  $\lambda$ and $\as$ values
    (``2-parton semi-anl.''). 
  }
  \label{fig:3-jet-anomdim-result}
\end{figure}

Our analysis has so far been for a $q\bar q$ event with a cloud of
soft gluons, and since we integrate over all $p_t$ values for the soft
gluons, it is in effect inclusive over the value of the observable.
Insofar as the linearity property of an observable holds beyond
$q\bar q$ events, we would expect the same modification factor as in
Eq.~(\ref{eq:Rsoln}) to hold for the $\Lambda/Q$ power correction also
in the $3$-jet limit.
We test this by starting from 3-parton events, adding a cloud of soft
gluons as done for Fig.~\ref{fig:all-order-many-lambdaf}, and then
inserting gluers into all dipoles and determining the change in the
value of the observable.
Fig.~\ref{fig:3-jet-anomdim-result} shows the resulting effective coefficient
of the power correction to the $C$-parameter, $\delta
C_\text{np}/[T_{q\bar q}/Q]$, as a function of the 
perturbative (3-parton) value of the $C$ parameter.
The red-dashed curve is without the cloud of soft gluons, 
the orange and cyan curves include the soft gluons.
The lower panel shows the ratio to the plain
$3$-parton (no-cloud) result for two finite $\as$ values at fixed $\lambda$.
The ratio coincides with the corresponding $2$-jet result at the same
$\lambda$ and finite $\as$ values, slightly different from
Eq.~(\ref{eq:Rsoln}) because of subleading logarithmic effects present
in the shower at finite $\as$ (``2-parton
semi-anl.''~\cite{supplement}, \S\ref{sec:supp:results-fixed-lambda}).
This confirms that the universality of the anomalous dimension holds
across the full spectrum.
We find similar results for the thrust.
Note that for small $V$, the $3$-parton result itself goes as
$c_V(1 - \frac12 \cS_1/\ln \frac1V)$, as required for consistency with
the anomalous dimension
\cite{supplement}~\S\ref{sec:supp:lit-comparisons} (see also
Ref.~\cite{Banfi:2025crj}).

Fig.~\ref{fig:3-jet-anomdim-result} uses a soft-gluon cloud with
$p_{t,\max} \ll Q$.
In practice, when the observable has a value $V\ll 1$, the actual
$p_{t,\max}$ is limited to be $\lesssim \xi V Q$, with $\xi$ of order
$1$.
Thus we expect the shift to involve a factor $\cR( \xi V Q)$ rather
than $\cR(Q)$, bringing additional $V$-dependence relative to the
shape of the pure 3-parton result. 
Note that for $VQ \sim \Lambda$, higher power corrections, including
full shape functions~\cite{Korchemsky:1999kt}, become relevant.

\begin{figure}
  \centering
  \includegraphics[width=\columnwidth]{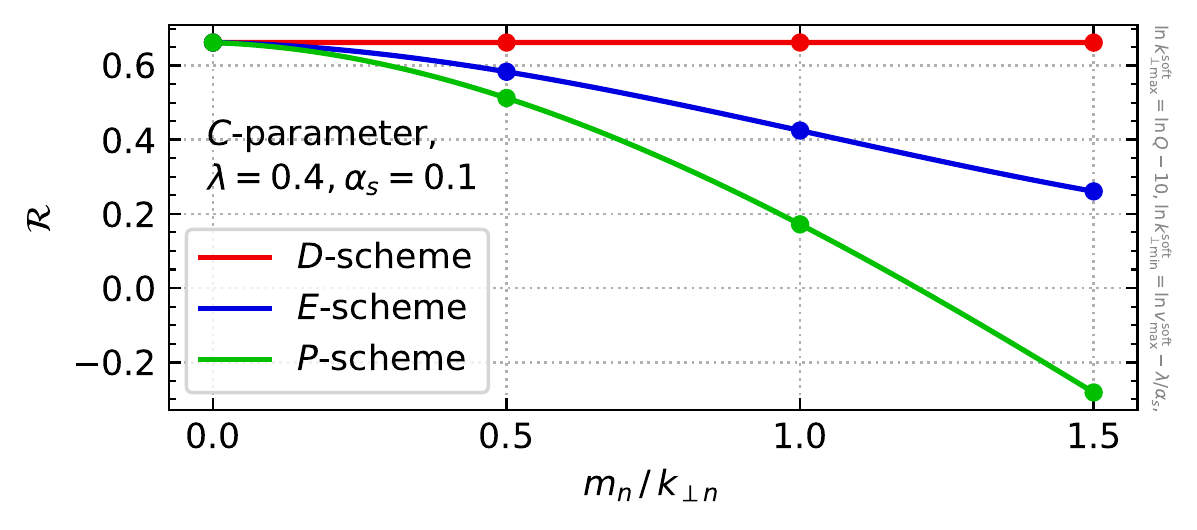}
  \caption{Dependence of the scaling factor $\cR$ on the ratio of
    gluer mass to transverse momentum, in three difference hadron-mass
    schemes.
    Only in the decay ($D$) scheme are the results consistent with
    Eq.~(\ref{eq:Rsoln}) also at non-zero gluer masses.  }
  \label{fig:mass-scheme}
\end{figure}

A further consideration is that hadrons have masses, while our
analysis so far has been for massless gluer emission.
There are various schemes for treating the hadron-mass dependence in
the definitions of observables~\cite{Salam:2001bd}.
Differences between schemes involve an anomalous
scaling~\cite{Salam:2001bd,Mateu:2012nk} that differs from
Eq.~(\ref{eq:Rsoln}) and whose normalisation involves non-perturbative
parameter(s) beyond $T_{q\bar q}$.
We conjecture that one can probe the effect of hadron masses by
examining the same class of shower kinematic maps with a
\emph{massive} gluer (\cite{supplement},
\S\ref{sec:supp:mass-schemes}).
The results, Fig.~\ref{fig:mass-scheme}, suggest one hadron-mass
scheme is favoured, the $D$ (or decay) scheme, where each hadron is
isotropically decayed to two massless particles.
In this scheme the anomalous dimension is universal, i.e.\ independent of
the ratio of the gluer mass $m_n$ to its transverse momentum $k_{\perp
  n}$.
In other schemes, $\cR$ depends strongly on $m_n/k_{\perp n}$,
implying that the hadronisation corrections depend on additional
non-perturbative parameters beyond $T_{q\bar q}$, connected with the
precise distribution of hadron masses and their transverse momenta.

\begin{figure}[t]
  \centering
  \includegraphics[width=\linewidth]{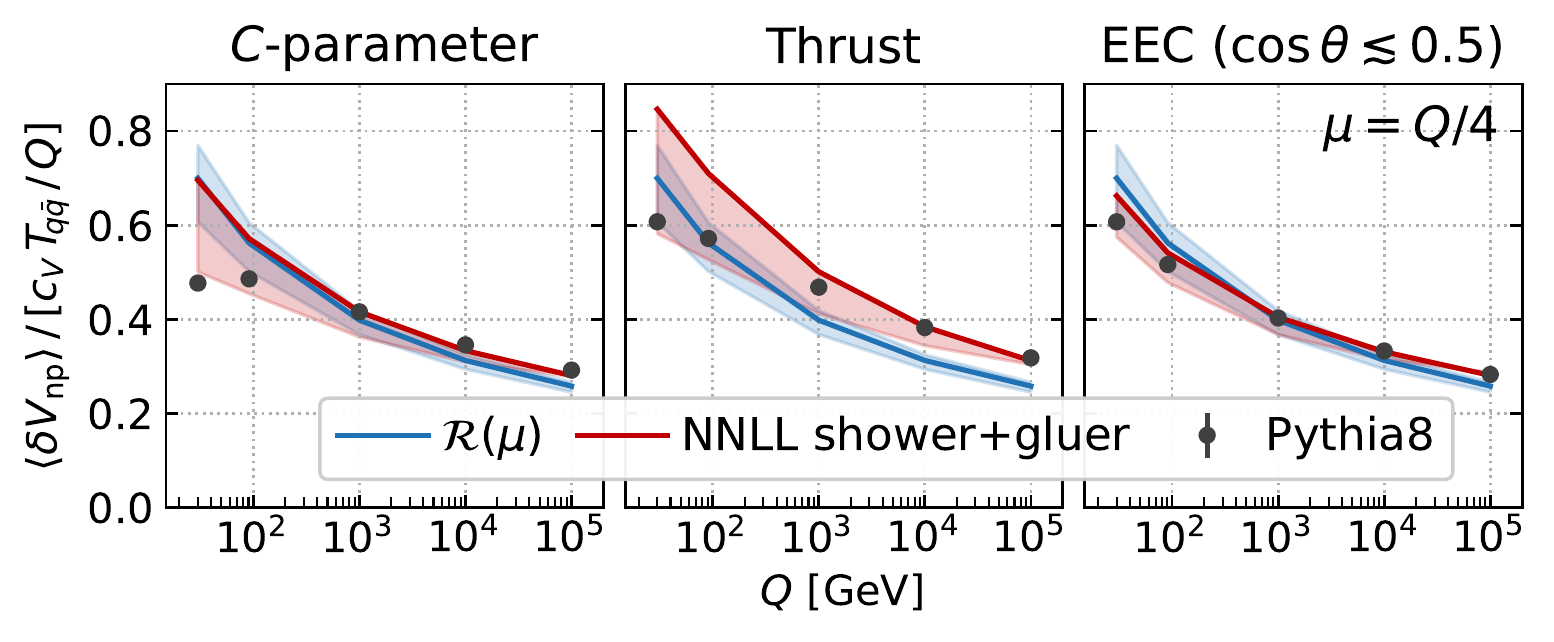}\\
  \includegraphics[width=\linewidth,page=2]{plotting/anomdim_v_fullshower_ktcut2.pdf}
  \caption{
    Top: average $C$-parameter, $\tau$ and EEC, as a function of $Q$,
    comparing the $R(\mu)$ calculation (Eq.~(\ref{eq:Rsoln})), gluer
    insertion into the NNLL PanGlobal shower, and Pythia8.
    For the EEC, $\lesssim0.5$ indicates a continuous bin
    boundary~\cite{supplement},\S~\ref{sec:supp:cont-bin}. 
    Bottom: the analogous set of results for the $C$-parameter
    distribution, with additionally the plain 3-parton result
    for comparison.
    In the grey area, we freeze $\mu$ in $\cR(\mu)$ to
    $\mu_\np = 2\GeV$.
  }
  \label{fig:pythia-comparison}
\end{figure}

We close with a comparison to hadronisation effects employing Pythia's
\cite{Bierlich:2022pfr} widely used Lund string
model~\cite{Andersson:1983ia,Andersson:1983jt}.
Exploring a range of $Q$ values,
Fig.~\ref{fig:pythia-comparison} (top) shows the coefficient of the
all-order non-perturbative correction to the average $C$-parameter,
thrust and EEC results, specifically the ratio to the plain $2$-parton
expectation. 
The blue curve is Eq.~(\ref{eq:deltaV-final}) with $\mu_\np = 2\GeV$,
and the band covers $1/6<\mu/Q<1/2$.
Our best prediction, the dark red curve, corresponds to the insertion
of an asymptotically soft gluer, $k_{\perp n}\simeq 0.0015\GeV$, into
actual showered events (NNLL PanGlobal~\cite{vanBeekveld:2024wws} with
a cutoff at $\mu_\np$).
The light red band shows the effect of increasing $k_{\perp n}$ to
$0.7\GeV$, which gives a sense of the magnitude of higher-power
effects.
These are surprisingly large even for $Q \sim 100\GeV$.
The difference between the dark red and blue curves gives a measure of
further higher-order perturbative effects.

The black points in Fig.~\ref{fig:pythia-comparison} show the
hadronisation corrections from Pythia 8.312
(Monash2013 tune \cite{Skands:2014pea}), using the $D$-scheme.
To relate those to Eq.~(\ref{eq:deltaV-final}), we need to determine
Pythia's effective $T_{q\bar q}$, which we do by requiring agreement
between the gluer insertion result and Pythia for the average
$C$-parameter at $Q=10^3\GeV$.
This yields $T_{q\bar q}^\text{Pythia} = 0.68\GeV$.
Across observables, there is good agreement between Pythia and our
predicted scaling, notably at high $Q$ values.

Finally, the lower panel of Fig.~\ref{fig:pythia-comparison} shows the full
$C$-parameter distribution at $Q=91.2\GeV$, with the same set of
curves as for the average, and additionally the pure $3$-parton result
for comparison (dashed-red).
The agreement between our curves and Pythia is striking and one
observes also a significant shape difference relative to the pure
$3$-parton result. This is connected with the dependence of the
anomalous dimension's scale on the $C$-parameter.
For other energies, and the thrust (subject to larger
subleading effects), see \cite{supplement},
\S\ref{sec:supp:more-distributions}.

To summarise: our main result is the determination of the all-order
structure of the anomalous scaling for the non-perturbative $\Lambda/Q$
power correction to widely-studied linear observables like the thrust,
$C$-parameter and energy correlators.
This has been done within a paradigm where recoil from gluer insertion
is linear in the gluer momentum~\cite{Caola:2021kzt,Caola:2022vea}.
While the paradigm remains to be fully established, our results are
sufficiently striking as to strongly motivate further study of its
theoretical foundations.
The most surprising finding is that despite the complicated structure
of the soft-gluon cloud from which a non-perturbative gluer is
emitted, the all-order result is a straightforward
exponentiation, Eq.~(\ref{eq:Rsoln}), of the first order
result~\cite{Dasgupta:2024znl}. 
This remains true even with a finite gluer mass, as long as one uses
the $D$-scheme.
Furthermore, this analytic approach reproduces the $Q$-dependence and
event-shape dependence seen in Pythia, to within
higher-order and higher-power effects that are larger than might have
been expected from earlier studies.
We look forward to future phenomenological applications of these
insights.\\

Note added: as this work was being finalised for publication,
Ref.~\cite{Chen:2026hmd} appeared, carrying out an operator-based
calculation for the anomalous scaling of the EEC hadronisation
correction in the $D$-scheme. Its result is consistent with ours,
providing further support for the validity of the underlying picture.\\
\vspace*{2em}

\begin{acknowledgments}
  We are grateful to Fabrizio Caola for collaboration in the early
  stages of this work and one of us (GPS) wishes to acknowledge Pier
  Monni for some initial joint work on the all-order corrections to
  the $C$-parameter hadronisation.
  We thank Paolo Nason and Giulia Zanderighi for providing the code
  from Ref.~\cite{Nason:2023asn} for the determination of the
  $3$-parton power corrections.
  We are grateful also to Mrinal Dasgupta for sharing the results of
  Ref.~\cite{Dasgupta:2024znl} ahead of publication.
  We wish to thank all of the above, as well as Melissa van Beekveld
  and Silvia Ferrario Ravasio, for numerous discussions and helpful
  comments on the manuscript.
  We are grateful also to Andrea Banfi and Basem Kamal El-Menoufi for
  discussions during the final stages of this paper about a
  forthcoming result of theirs~\cite{Banfi:2025crj} that
  coincides with our Eq.~(\ref{eq:S1-and-lambda}) and with findings of our
  \cite{supplement}, \S\ref{sec:supp:lit-comparisons}.
  This work has been funded by the European Research Council (ERC)
  under the European Union's Horizon 2020 research and innovation
  programme (grant agreement No.\ 788223, JH, GPS, SZ),
  by a Royal Society Research Professorship
  (RP$\backslash$R1$\backslash$231001, GPS),
  by the Science and Technology Facilities Council (STFC) under
  grants ST/T000864/1 (GPS), ST/X000761/1 (GPS, SZ)
  and by the Australian Research Council via Discovery Project DP220103512 (JH).
\end{acknowledgments}

\bibliographystyle{apsrev4-2}
\bibliography{MC}

\input{supplementary_material}

\end{document}

%% file: supplementary_material.tex
\newpage
\onecolumngrid
\newpage


\appendix

\section*{Supplemental material}

\subsection{Comparison of PanGlobal and PanLocal results to those in
  the literature}
\label{sec:supp:lit-comparisons}

The core formula that we use when identifying the coefficient of the
non-perturbative power correction with a given kinematic map is
\begin{subequations}
  \label{eq:core}
  \begin{align}
    \langle \delta V_\np\rangle_{q\bar q}
    &=
      \int_0^{\mu_\np} \frac{dk_{tn}}{k_{tn}} d\eta_n \frac{2C_F
      \as^{(\text{eff})}(k_{tn})}{\pi}
      \left[V(p_1,p_2,k_n) - V(\tilde p_1,\tilde p_2)\right],
    \\
    &=
      \label{eq:core-with-lim}
      \frac{T_{q\bar q}}{Q}    
      \lim_{\mu_0\to 0}
      \int_0^{\mu_\np} \frac{dk_{tn}}{k_{tn}}\, d\eta_n\,
      Q \delta(k_{t n}-\mu_0)
      \times \left[V(p_1,p_2,k_n) - V(\tilde p_1,\tilde p_2)\right]
      + \order{\frac{\Lambda^2}{Q^2}},
  \end{align}
\end{subequations}
where the linearity properties of the observable ensure that the limit
is well defined.

In our study we employ the kinematic maps of two different parton
showers, namely PanGlobal (PG) and PanLocal (PL), both with
transverse-momentum ordering, i.e. $\beta_{\text{PS}} = 0$ in the
language of Ref.~\cite{Dasgupta:2020fwr}.\footnote{Given that the
  PanLocal shower requires $0 < \beta_{\text{PS}} <1$ for logarithmic
  accuracy, it may seem surprising that we use it here with
  $\beta_{\text{PS}} = 0$.
  The point is that we are probing the effect of a gluer insertion
  with an asymptotically soft gluer, $\mu_0 \to 0$.
  Thus, the choice of the ordering variable makes no difference when probing the
  effect of the soft gluer, because that soft gluer will always come as the
  last step regardless of $\beta_{\text{PS}}$.
  Note that for linear observables, the transverse recoil contributions,
  which would cause problems with logarithmic accuracy with
  $\beta_{\text{PS}}=0$, simply average to zero after azimuthal
  integration when evaluating
  gluer insertion effects.
}
The full details of the
kinematic maps can be found in Appendix~A of
Ref.~\cite{vanBeekveld:2025lpz}, however the essence is as follows.
For a dipole $ij$, we define
\begin{equation}
\widetilde{s}_{ij} = 2 \tilde{p}_i \cdot \tilde{p}_j\,, \quad \widetilde{s}_{i}
	= 2 \tilde{p}_i \cdot Q\,, \quad \widetilde{s}_{j} = 2 \tilde{p}_j \cdot Q\,,
\end{equation}
where $Q$ is momentum of the $e^+e^-$ system, and introduce
\begin{equation}
z_+ = k_{\perp n} \sqrt{\frac{ \tilde{s}_j}{{\tilde s}_{ij}\tilde{s}_i}} e^{+\bar{\eta}_n},
\qquad
z_- = k_{\perp n} \sqrt{\frac{ \tilde{s}_i}{{\tilde s}_{ij}\tilde{s}_j}} e^{-\bar{\eta}_n},
\label{eq:alphakbetak}
\end{equation}
where $k_{\perp n}$ is the transverse momentum with respect to the
$ij$ dipole, as opposed to $k_t$, which is the transverse momentum with
respect to the original $q\bar q$ direction.
For a $q\bar q$ system they are the same, but for a general dipole
they differ.
For the PanGlobal map, we have
\begin{subequations}
\label{eq:panglobal-map}
\begin{align}
\bar k_n^{\mu}  &= r_L( z_+ \tilde{p}_i^{\mu}  + z_- \tilde{p}_j^{\mu}  + k_{\perp n}^{\mu} ), \\
\bar p_i^{\mu}  &= r_L(1 - z_+)\tilde{p}_i ^{\mu} , \\
\bar p_j ^{\mu} &= r_L(1 - z_-)\tilde{p}_j^{\mu},
\end{align}
\end{subequations}
where $k_{n}^{\mu}$ is the momentum of the gluer, with $(k_{\perp n}^\mu)^2
= -(k_{\perp n})^2$. The gluer's transverse momentum with respect to the
parent dipole, $|k_{\perp n}|$, satisfies $2 z_+z_-
\ptilde_i.\ptilde_j = |k_{\perp n}|^2$.
The rescaling factor $r_L$ is chosen to ensure
conservation of the squared 4-momentum of the system.
All the event momenta then undergo a common boost to bring the
event back to its original centre of mass.
The details~\cite{FerrarioRavasio:2023kyg} of the rescaling and global
boost only affect quadratic corrections.

For the PanLocal map, we have
\begin{subequations}
\label{eq:panlocal-map}
\begin{align}
k_n^{\mu}  &= z_+ \tilde{p}_i^{\mu}  + z_- \tilde{p}_j^{\mu}  + k_{\perp n}^{\mu} , \\
p_i^{\mu}  &=(1 - z_+)\tilde{p}_i ^{\mu} +
                   \frac{z_+z_-}{1-z_+}
                   \tilde{p}_j^{\mu} -  k_{\perp n}^{\mu}, \\
p_j ^{\mu} &=\frac{1 - z_+ - z_-}{1 - z_+}\tilde{p}_j^{\mu},
\end{align}
\end{subequations}
which explicitly conserves momentum, so no further boosts are
required.
The equations here are given for the case where $i$ is the
``splitter'', i.e.\ the parent dipole particle that absorbs the
transverse component.
In the PanLocal map, one chooses $i$ to be the splitter with probability $f(\bar \eta_n)$
and $j$ with probability $1-f(\bar \eta_n)$, with $f(\bar \eta)$ given
by
\begin{equation}
  f(\bar{\eta}) =
  \begin{cases}
    0 & \text{if } \bar{\eta} <-1 \\
    \frac{15}{16}\left(\frac{\bar{\eta}^5}{5}-\frac{2\bar{\eta}^3}{3}+\bar{\eta}
    +\frac{8}{15} \right) & \mbox{if } -1\leq \bar{\eta} \leq 1 \\
    1 & \text{if } \bar\eta>1
  \end{cases}.
  \label{eq:fdip}
\end{equation}
Based on the discussion in Refs.~\cite{Caola:2021kzt,Caola:2022vea},
both of these maps are expected to be suitable for determining the
coefficient of the linear power correction, with the critical element
being the continuity in the treatment of the recoil momenta as a
function of $\bar \eta_n$.
One should be aware that there remain open questions as to the
validity of this general approach when one of the dipole ends is a
gluon.
In the large-$N_c$ limit, given that we are considering just soft
physics, one could make the argument that it is plausible that gluons
and quarks should be treated similarly.
Existing results in
Refs.~\cite{Caola:2021kzt,Caola:2022vea,Nason:2023asn,Nason:2025qbx}
also use such an assumption.
A further point to keep in mind is that we consider the emission of
just a single soft on-shell gluer, rather than an off-shell gluer that
splits to a gluon pair or $q\bar q$ pair.
For observables linear in the soft limit, such as those that we
consider here, the single on-shell gluer approach reproduces the
structure of a full off-shell approach to within a universal
``Milan-factor'' constant~\cite{Dokshitzer:1997iz,Dokshitzer:1998pt,Dasgupta:1998xt,Dasgupta:1999mb}.

We now show that our approach correctly reproduces results already
available in the literature. We start by calculations 
the leading non-perturbative power correction for
$3$-parton configurations, and comparing to the results of \cite{Caola:2021kzt,Caola:2022vea} as
embodied in the code provided by the authors of Ref.~\cite{Nason:2023asn}.
We generate a $q\bar q g$ event using both PG and PL as tree-level
generator, and we include the correct $q\bar q g$ matrix element using the
multiplicative matching of Ref.~\cite{Hamilton:2023dwb}.
The gluon transverse momentum $p_{tg}$ is constrained to be larger
than the shower cutoff $p_{t, \text{min}}$, $p_{tg} > p_{t,\text{min}}$.
Given this constraint, the kinematics of the perturbative gluon is sampled
randomly by the parton shower. 
Starting from this 3-parton configuration, a non-perturbative gluer is
emitted with a transverse momentum $k_{\perp n}$ much smaller
than the shower cutoff, $k_{\perp n} \ll p_{t,\text{min}}$.
\begin{figure}[tb]
  \centering
  \includegraphics[width=0.45\textwidth,page=1]{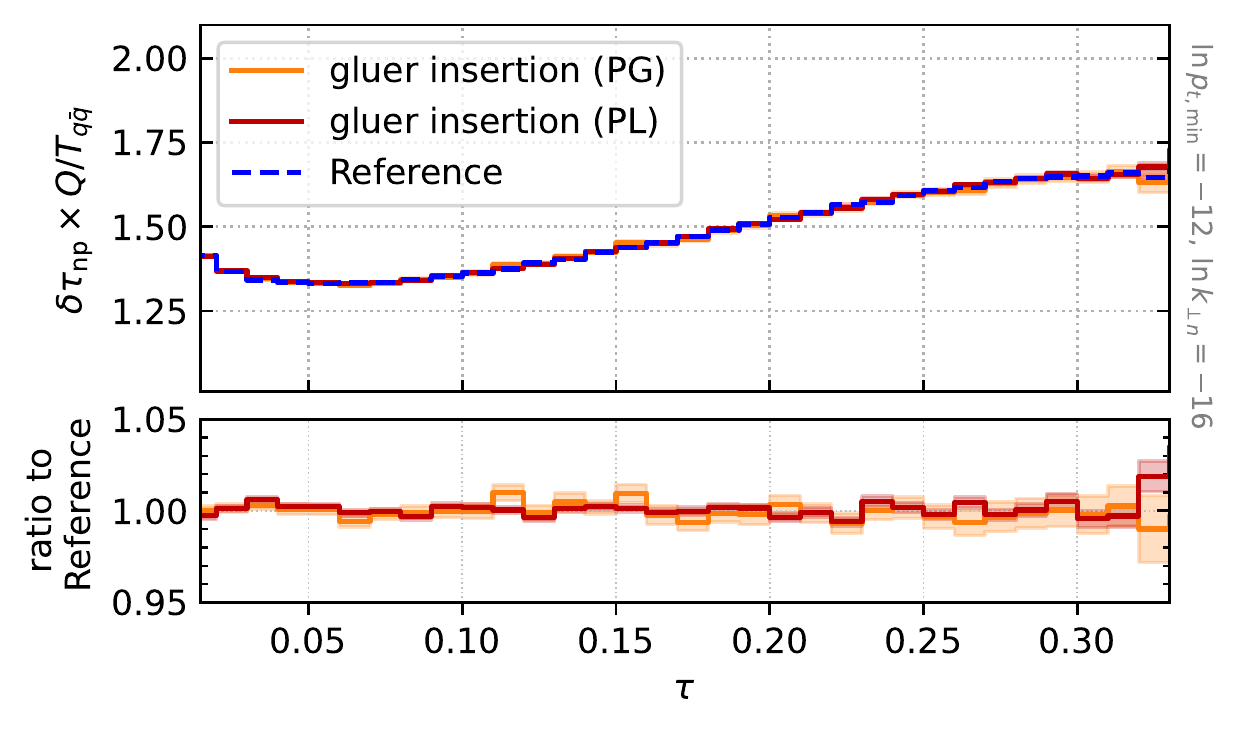}
  \includegraphics[width=0.45\textwidth,page=2]{plotting/comparison-giulia-paolo/fo_comparison_giulia_paolo.pdf}
  \caption{Average non-perturbative shift for thrust (left) and
    $C$-parameter (right) for 3-parton configurations.
    Results are obtained for PanGlobal (orange, solid)
    and PanLocal (dark red, solid), and compared to the Reference results of \cite{Caola:2021kzt,Caola:2022vea} as
    embodied in the code provided by the authors of
    Ref.~\cite{Nason:2023asn} (blue, dashed), in the large-N$_C$ limit. }
  \label{fig:comparison-GP}
\end{figure}

Our results are obtained in the large-$N_c$ limit, as standard for
dipole showers, while Refs.~\cite{Caola:2021kzt,Caola:2022vea,Nason:2023asn} perform the
calculation in full colour.  In order to perform this comparison, we
thus extract the large-$N_c$ limit of Ref.~\cite{Nason:2023asn}
considering only contributions from the $qg$ and $\bar q g$ dipoles,
which are proportional to $\CA/2$, and neglecting the $q\bar q$
dipole, as it would generate terms proportional to $C_F-\CA/2$, which
vanish in limit we are considering.
Fig.~\ref{fig:comparison-GP} shows the leading non-perturbative
shift in a $3$-jet configuration for the thrust (left) and the
$C$-parameter (right) as obtained with PanGlobal and PanLocal, compared to
Refs.~\cite{Caola:2021kzt,Caola:2022vea,Nason:2023asn} (``Reference''). In both cases we find
agreement with the reference results.

We conclude this section by reproducing the value of the anomalous
dimension $\mathcal{S}_1$ through a numerical study of the behaviour of the
non-perturbative power correction in the approach to the $2$-jet
limit.
In particular, we connect to the result of \cite{Dasgupta:2024znl},
where $\mathcal{S}_{1}$ has been obtained in a semi-analytic way
for the thrust and the $C$-parameter.
We generate a $q\bar q$ event and emit a soft perturbative gluon at a
scale higher than the shower cutoff, $k_{tg}>k_{t, \text{min}}$. The
differential weight associated to this configuration is $d\sigma$.
The non-perturbative correction is then probed through the emission of
a very soft gluer, $k_{tn}\ll k_{t, \text{min}}$.

The average non-perturbative shift to an event-shape $V$ at NLO
accuracy is obtained as
\begin{equation}
  \label{eq:plateau}
  \langle\delta V\rangle^\text{NLO}_{\text{np}}=
  \int d\ln V \,\frac{d\langle\delta
    V\rangle^\text{NLO}_{\text{np}}}{d\ln V}
  = 
  \int d\ln{V}\big({\langle\delta V\rangle}_{\text{np}}  - {\langle\delta V\rangle}_{\text{np}}^{\text{Born}} \big)\, \frac{1}{\sigma}\frac{d\sigma}{d\ln V}\,,
\end{equation}
where the term in brackets represents the difference between the average
non-perturbative shift in a $3$-parton configuration ($q\bar q g$) and in
a Born ($q\bar q$) configuration.
This expression has been obtained as follows: at NLO accuracy, we need
to consider $\mathcal{O}(\alpha_s)$ corrections associated with the
emission of a perturbative gluon $g$ from a $q\bar q$ system.
When considering the leading anomalous dimension $\mathcal{S}_1$, we
work in the asymptotic region, where real and virtual corrections are
equal but with opposite sign.
Real and virtual corrections are summed together: when integrating
over the gluer insertion, we obtain the term in brackets, which is the
difference of the contribution given by a gluer insertion in a $q\bar
q g$ event (real correction) and in a $q\bar q $ event (virtual
correction). This expression is reweighted by the probability of
emitting a soft perturbative gluon $d\sigma$, which is obtained
integrating over the gluon kinematics.
Note that the term in brackets is expected to have an inverse
logarithmic scaling in the event 
shape $V$ which is governed by the anomalous dimension
$\mathcal{S}_1$: 
\logbook{}{see
  panscales-pc-clean/plotting/approach-2jet-limit/approach-2jet-extra.pdf
for checks of coefficients for this and the next equation}
\begin{equation}
  \label{eq:plateau-2}
  {\langle\delta V\rangle}_{\text{np}}  - {\langle\delta
    V\rangle}_{\text{np}}^{\text{Born}} =
  -
  \frac{\CA}{4C_F}
  \frac{\mathcal{S}_1}{(\ln d_V/V) - 3/4}
  \frac{c_V T_{q\bar q}}{Q}
  \,,
  \qquad
  |\ln V| \gg 1\,,
  \qquad
  \left\{
    \begin{array}{ll}
      d_C   \!\!\! &= 6\\
      d_\tau\!\!\! &= 1
    \end{array}
  \right..
\end{equation}
The potential presence of a $1/\ln V$ structure had already been noted
in Ref.~\cite{Nason:2023asn}.
The inverse logarithmic scaling arises because it is only when the
perturbative gluon is at central rapidities that $\langle \delta
V\rangle_\np$ differs substantially from $\langle \delta
V\rangle_\np^\text{Born}$, cf.\ Fig.~\ref{fig:one-emission}. 
That the normalisation should be as written is a consequence of the
following argument.
When approaching the $2$-jet limit ($V \ll 1$), the perturbative
distributions for the thrust and $C$-parameter have a well-known
logarithmic enhancement 
\begin{equation}
  \frac{1}{\sigma}\frac{d\sigma}{d\ln 1/V}  = \frac{2 \as C_F}{\pi} \left(\ln{\frac{d_V}{V}}-\frac34\right)\,,
  \qquad |\ln V| \gg 1\,.
\end{equation}
Thus the integrand in Eq.~\eqref{eq:plateau} is free of logarithms and
tends to a constant for large $|\ln V|$.
The integral must reproduce the anomalous dimension shown in the text,
i.e.\ the integrand must tend to $-c_V
\frac{\CA\alpha_s}{2\pi}\mathcal{S}_1 \cdot T_{q\bar q}/Q$, and
this tells us the expected normalisation in Eq.~(\ref{eq:plateau-2}). 

\begin{figure}[tb]
  \centering
  \includegraphics[width=0.4\textwidth,page=1]{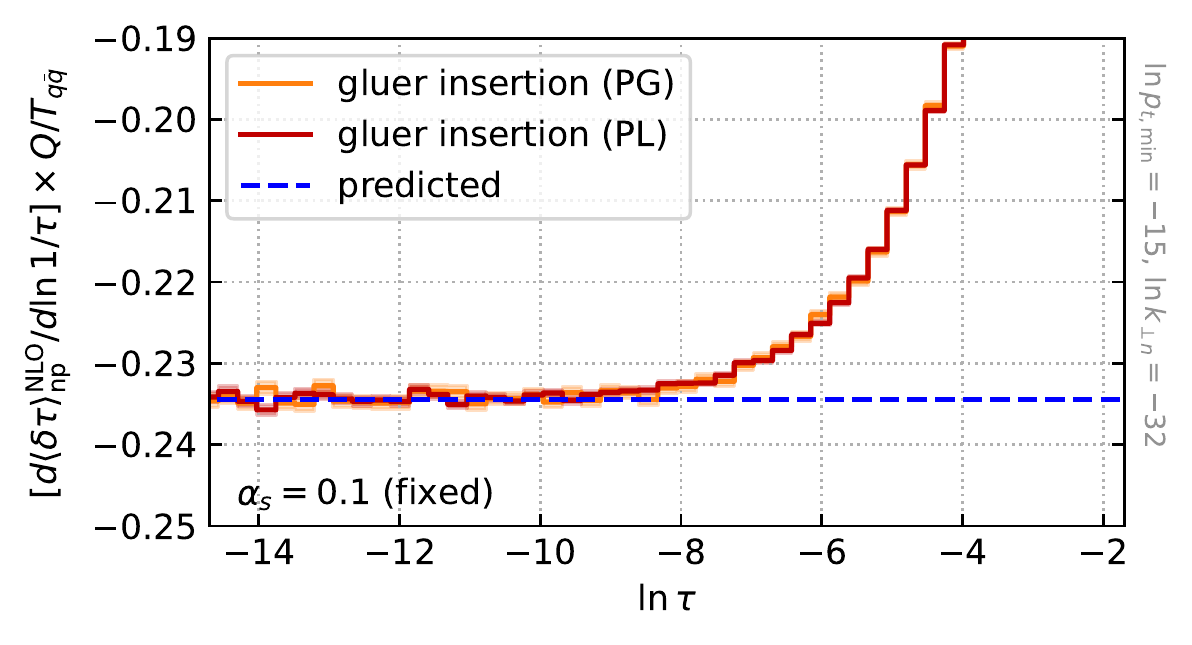}
  \includegraphics[width=0.4\textwidth,page=2]{plotting/approach-2jet-limit/approach-2jet.pdf}
  \caption{Approach to the $2$-jet limit for thrust (left) and
    $C$-parameter (right), showing the integrand of Eq.~(\ref{eq:plateau}).
    Results are given for the PanGlobal (orange, solid)
    and PanLocal (dark red, solid) maps.
    The predicted value (blue, dotted) corresponds to $-c_V \frac{\CA\alpha_s}{2\pi}\mathcal{S}_1$.
  }
  \label{fig:comparison-DH}
\end{figure}

Figure \ref{fig:comparison-DH} shows the numerically approach of the
integrand to the $2$-jet limit for the thrust (left) and the
$C$-parameter (right), using the PanGlobal (orange, solid) and PanLocal
(dark red, solid) maps for gluer insertion.
In both cases, as $|\ln V|$ becomes large, the results tend to a
plateau, which is agreement with the predicted value.

\subsection{Analytic derivation of $\cS_1$}
\label{sec:supp:S1}

\logbook{8a26a86}{See ../maths/thrust-anom-dim.nb}

The essence of our derivation will be to consider an observable that
is the sum of transverse momenta inside a rapidity window
$|\eta|< Y$ with $Y\gg 1$\,
\begin{equation}
  \label{eq:SY-def}
  S_t(Y) = \sum_i p_{ti} \Theta(|\eta_i| < Y)\,,
\end{equation}
where the $p_i$ in this definition are to be understood as running
over all momenta, whether perturbative or non-perturbative.
One may define the $p_{ti}$ to be the transverse component with respect to the thrust axis or with
respect to some specific jet-axis definition for each of two exclusive
jets.
We will consider $Y$ to be large but not so large that it includes the
$q$ or $\qbar$.

The key quantity that we want to calculate is
the change in $S_t(Y)$ due to gluer emission. 
In a $q\bar q$ event, this is simply given by
\begin{equation}
  \label{eq:6}
  \langle \delta S_t(Y) \rangle_{q\bar q} = 2Y \, T_{q\bar q}\,.
\end{equation}

Next, we consider a $q g \bar q$ event, where a soft perturbative gluon
$g$ has been emitted at rapidity $\eta_g=0$. Aligning the $q\bar q$
system along the $z$-direction, the gluon $4$-momentum
($p_x,p_y,p_z,E$) is
\begin{equation}
  \label{eq:anlcS1-ptildeg}
  \tilde p_g = \tilde p_{tg}(0,1,0,1)\,,
\end{equation}
where the transverse momentum with respect to the $q\bar q$ axis
($\tilde p_{tg}$) is much smaller than the hard scale $Q$,
$\tilde p_{t,g} \ll Q$.
We now insert a non-perturbative gluer $k_n$ into the event.
In the the main text and the numerical studies, we discussed
$T_{qg\bar q}(\eta,\eta_g) \equiv \rho_{qg\bar q}(\eta,\eta_g)
T_{q\bar q}$, as an integral over all gluer kinematics with gluer
rapidity $\eta_n$ equal to $\eta$.
For an analytical calculation this is not the most convenient
approach, and instead we will bypass $\rho_{qg\bar q}(\eta,\eta_g)$,
taking a path that in effect allows us to obtain result for the integral
$\int_{-\infty}^{+\infty} d\eta (\rho_{qg\bar q}(\eta,\eta_g)-1)$.
That is equivalent to Eq.~(\ref{eq:R-integral}) once one takes into
account that $\rho_{qg\bar q}(\eta,\eta_g)$ is a function only of
$\eta - \eta_g$, a consequence of the longitudinal boost invariance of
the maps.

We focus first on the $\bar q g$ dipole (where the $\bar q$ has
negative $p_z$) and apply the map in
Eq.~(\ref{eq:panglobal-map}) with $i=g$, $j=\qbar$
($p_{\qbar z} < 0$).
After application of the map, the momenta of interest are those that
are potentially inside the rapidity window, i.e.\ $p_g$ and $k_n$
(using $r_L \to 1$),
\logbook{}{pnEta0ynRep in thrust-anom-dim.nb}
\begin{subequations}
  \begin{align}
    p_g &= (1-z_+) \ptilde_{tg}(0,1,0,1)\,,
    \\
    k_n &= \left(
          k_{\perp n} \sin \phi_n ,\;
          \ptilde_{tg} z_+ + k_{\perp n} \cos \phi_n ,\;
          - \frac{k_{\perp n}^2}{2 \ptilde_{tg} z_+} - k_{\perp n} \cos \phi_n
          ,\;
          \frac{k_{\perp n}^2}{2\ptilde_{tg} z_+}
          + \ptilde_{tg} z_+
          +  k_{\perp n}  \cos\phi_n
          \right)\,.
  \end{align}
\end{subequations}
The perturbative gluon itself will always be in the window.
To determine whether the non-perturbative gluer will be in the window,
we consider the limit $z_+ \ll k_{\perp n}/\ptilde_{tg} $ , where we obtain
\begin{equation}
  \label{eq:yn-gluer}
  \eta_n
  = -\log \frac{k_{\perp n}}{\ptilde_{tg} z_+} > -Y
  \quad \to \quad
  z_+ > \frac{k_{\perp n}}{\ptilde_{tg}} e^{-Y}\,.
\end{equation}
Thus we have
\begin{equation}
  \label{eq:deltaSY}
  \delta S_t(Y) = -\ptilde_{tg} z_+ +
  \sqrt{k_{\perp n}
    + \ptilde_{tg}^2 z_+^2
    + 2 z_+ \ptilde_{tg} k_{\perp n} \cos\phi}\,
  \Theta\left(z_+ > \frac{k_{\perp n}}{\ptilde_{tg}} e^{-Y}\right)\,.
\end{equation}
The integral over the gluer kinematics so as to obtain the average
shift can be written as
\begin{equation}
  \label{eq:avg-deltaSY} 
  \langle \delta S_t(Y)\rangle_{\qbar g}
  =
  T_{q\bar q} \lim_{k_{\perp n} \to 0}
  \left[
    \frac{1}{k_{\perp n}}
    \int_0^{2\pi} \frac{d\phi}{2\pi}
    \int_{\frac{k_{\perp n}}{\ptilde_{tg}} e^{-Y}}^1 \frac{dz_+}{z_+} \,
    \delta S_t(Y)
  \right].
\end{equation}

For the time being we work in the large-$N_C$ limit, so that the $\bar
q g$ dipole has the same colour factor as the $q\bar q$ dipole,
allowing us to reuse $T_{q\bar q}$.
Additionally, in setting the lower limit on the $z_+$ integral, we have
exploited the fact that the $-\ptilde_{tg} z_+$ term in
Eq.~(\ref{eq:deltaSY}) contributes negligibly for $z_+$ values below the
limit, yielding a term suppressed by a power of $e^{-Y}$, which we can
omit.
To evaluate Eq.~(\ref{eq:avg-deltaSY}) we first carry out the $z_+$
integration,
\begin{equation}
  \label{eq:avg-deltaSY-step2} 
  \frac{\langle \delta S_t(Y)\rangle_{\qbar g}}{T_{q\bar q}}
  =
  \lim_{k_{\perp n} \to 0}
  \int_0^{2\pi} \frac{d\phi}{2\pi}
  \bigg[
    F(z_+,\phi)
  \bigg]_{z_+=\frac{k_{\perp n}}{\ptilde_{tg}} e^{-Y}}^{z_+=1}
\end{equation}
with
\begin{multline}
  F(z_+,\phi) = 
    \frac{\sqrt{\ptilde_{tg}^2 z_+^2+2 \ptilde_{tg} k_{\perp n} z_+ \cos
        (\phi )+k_{\perp n}^2}}{k_{\perp n}}
    -\frac{\ptilde_{tg} z_+}{k_{\perp n}}
    -\tanh ^{-1}
    \left(
      \frac{\ptilde_{tg} z_+ \cos (\phi
        )+k_{\perp n}}{\sqrt{\ptilde_{tg}^2 z_+^2+2 \ptilde_{tg}
          k_{\perp n} z_+ \cos (\phi )+k_{\perp n}^2}}
    \right)
    +\\
    +\cos (\phi ) \tanh ^{-1}\left(\frac{\ptilde_{tg}      
        z_+ +k_{\perp n} \cos (\phi )}{\sqrt{\ptilde_{tg}^2 z_+^2+2
          \ptilde_{tg} k_{\perp n} z_+ \cos (\phi
          )
        +k_{\perp n}^2}}\right).
\end{multline}
With some manipulation, one can show that
\begin{equation}
  \label{eq:F1-twophivals}
  F(1,\phi) +  F(1,\phi+\pi) = \order{\frac{k_{\perp n}}{\ptilde_{tg}}}\,,
\end{equation}
such that the upper $z_+$ limit in Eq.~(\ref{eq:avg-deltaSY-step2})
vanishes. 
As concerns the lower limit, one can show that 
\begin{equation}
  \label{eq:FsmallZ}
  F\left(
    \frac{k_{\perp n}}{\ptilde_{tg}} e^{-Y},\phi
  \right) =
  -Y
  - \frac{1}{2} \log \left(4 \csc ^2\phi \right)
  -\frac{1}{2} \cos (\phi ) \log \left(\tan^2 \frac{\phi}{2}\right)
  +1
  + \order{e^{-Y}}\,.
\end{equation}
Taking the limit of large $Y$ and small
${k_{\perp n}}/{\ptilde_{tg}}$, one can then carry out the $\phi$
integration, obtaining
\begin{equation}
  \label{eq:avg-deltaSY-result} 
  \frac{\langle \delta S_t(Y)\rangle_{\qbar g}}{T_{q\bar q}}
  =
  Y - 2(1 - \log 2)\,.
\end{equation}
Taking into account the second ($gq$) dipole and evaluating the
difference between the $\qbar g + gq$ and the $\qbar q$ contributions
to $\delta S_t(Y)$ gives us
\begin{equation}
  \label{eq:sum-of-SYs}
  \langle \delta S_t(Y)\rangle_{\qbar g}
  +
  \langle \delta S_t(Y)\rangle_{g q}
  -
  \langle \delta S_t(Y)\rangle_{\qbar q}
  = - 4(1-\log 2)T_{q\bar q}
\end{equation}
which corresponds to the result given in Eq.~(\ref{eq:R-integral}),
but derived bypassing the intermediate $\rho_{qg\qbar}(\eta,\eta_g)$
quantity, which would have been more complicated to handle
analytically.

We close this section with a brief comment on the subleading-colour
contributions.
At full colour, there would be a $\CA/(2C_F)$ factor in
$\langle \delta S_t(Y)\rangle_{\qbar g}$ and
$\langle \delta S_t(Y)\rangle_{g q}$, and an additional
$(2C_F - \CA)/(2C_F)\langle \delta S_t(Y)\rangle_{\qbar q}$ dipole
contribution.
Taken together this would give an overall $\CA/2C_F$ factor in the
full-colour equivalent of Eq.~(\ref{eq:sum-of-SYs}).
That $\CA/2C_F$ factor would multiply the $C_F$ factor in
Eq.~(\ref{eq:RNLO-step}) ultimately giving us the $\CA$ factor as in
Eq.~(\ref{eq:RNLO-final}).

\subsection{Universality of longitudinal recoil}
\label{sec:supp:perp-v-long}

The observables that we consider in the main text all have the
property that $f_V(\eta)$ in Eq.~(\ref{eq:V-linearity}) vanishes for
large soft-particle rapidities $\eta_i$, so that for a $q\bar q$
system longitudinal recoil of the $q$ and $\qbar$ does not affect the
observable.
However, as soon as we have a perturbative gluon at large angles,
emission of a gluer from a $qg$ dipole will induce longitudinal recoil
of the perturbative gluon, where longitudinal is now defined with
respect to the $qg$ dipole directions.
Since the perturbative gluon is at large angles, this will affect the
observable.
A key assumption that underlies our derivation of
Eq.~(\ref{eq:dRdlnQ}) is that if we have a perturbative $qg\qbar$
system, then in any small patch of phase space that contains the
perturbative gluon, the all-order resummation for the power
correction to the energy flow in the patch will have the same
correction factor as a patch without a gluon.
The purpose of this section is to provide an understanding of why this
condition is true.

The route that we take to establish this point is to start with a
$q\bar q$ system.
We will demonstrate that the corrections to energy flow in a patch
containing the quark or anti-quark acquire the same first order
correction as a patch not containing either.
Once this is established, then for a $qg\qbar$ system the same
argument will hold as concerns energy flow in a patch containing the
$g$ end of each of the $qg$ and $g\qbar$ dipoles with respect to
emission of a further gluon from each of those dipoles.
Recursively applying this argument gives the all-order demonstration
of the property that we need.

We start with a $q\qbar$ system and consider the light-cone plus
component at rapidities larger than $Y$
\begin{equation}
  \label{eq:Splus-Y}
  S_+(Y) = \sum_i p_{+i} \Theta(\eta_i > Y)\,,
\end{equation}
where $p_{+i} = E_i+p_{zi}$ and the sum includes the $q$ and $\qbar$
if they are in the rapidity region (as in Eq.~(\ref{eq:SY-def}), the
$p_i$ run over all momenta, whether perturbative or non-perturbative).
One way of analysing the observable is to note that if we take the limit
$Y \to -\infty$, then $S_+(-\infty) = Q$, since all the $z$
components cancel and the sum of energy components must add up to the
centre-of-mass energy.
Thus $S_+(-\infty)$ has no power correction, i.e.\
$\langle \delta S_+(-\infty) \rangle_{\qbar q}=0$.
Therefore we can express $\langle\delta S_+(Y)\rangle$ by considering
the difference between $S_+(Y)$ and $S_+(-\infty)$, which involves a
convergent integral over rapidities,
\begin{equation}
  \label{eq:Splus-from-Tqq}
  \langle \delta S_+(Y) \rangle_{q\qbar} 
  =
  \langle \delta S_+(-\infty) \rangle_{q\qbar} 
  - T_{q\qbar} \int_{-\infty}^Y
  d\eta \, e^{\eta}
  = -T_{q\qbar}\, e^{Y}.
\end{equation}
Next, we consider a system with a soft perturbative gluon.
For a specific soft-gluon rapidity $\eta_g$, the impact of the gluer on
transverse momentum flow has a strong dependence on the rapidity,
cf.\ Fig.~\ref{fig:one-emission}.
Similarly the impact of the gluer on $S_+(Y)$ will depend strongly on
which dipole we consider and whether the perturbative gluon has
rapidity larger or smaller than $Y$ (e.g.\ if the $\eta_g \ll Y$ and one
considers the $qg$ dipole with the quark having positive $p_z$, then
$\delta S_+(Y)_{qg} $ will be the same as $\delta S_+(Y)_{q\qbar}$,
while $\delta S_+(Y)_{\qbar g}$ will be zero).
However, once we integrate over the rapidity of the perturbative soft
gluon, the impact of the gluer on the transverse momentum per unit
rapidity will be independent of rapidity.
Let us label the net non-perturbative transverse momentum per unit
rapidity, after integrating over the perturbative gluon, as
$T_{q g \bar q}$.
Since $S_+(-\infty)$ always has the property that its non-perturbative
change is zero, we can follow the same logic as used in
Eq.~(\ref{eq:Splus-from-Tqq}) to obtain
\begin{equation}
  \label{eq:Splus-from-Tqgq}
  \langle \delta S_+(Y) \rangle_{qg \bar q} 
  =
  -T_{qg\qbar}\, e^{Y}.
\end{equation}
This same argument can be applied to any order as long as the effect
of the gluer, after averaging over perturbative emissions, remains
independent of rapidity.


As a further verification, we also carried out an explicit numerical
study.
Rather than a sharp boundary $\Theta(\eta_i>Y)$, numerically it is more
stable to use a boundary with a continuous turn-on, cf.\ also \S\ref{sec:supp:cont-bin}.
Specifically, we include a particle's plus component with a weight
$f_{0\pm\tfrac12}(\eta)$ corresponding to a continuous turn on between
$\eta=-\tfrac12$ and $=+\tfrac12$
\begin{equation}
  \label{eq:fuzzy-cone}
  \tilde S_+ = \sum_i p_{+i}\, f_{0\pm\tfrac12}(\eta_i)\,,
  \qquad
    f_{0\pm\tfrac12}(\eta) = \Theta\left(-\frac12 < \eta <
    \frac12\right)\left(\eta+\frac12\right)  + \Theta\left(\eta \ge \frac12 \right)\,.
\end{equation}
We then look at how $\tilde S_+$ changes
when integrating over all possible gluer insertions.
For a $q\bar q$ system, it is straightforward to show that this leads
to the relation 
\begin{equation}
  \label{eq:fuzzy-cone-expectation}
  \langle \delta \tilde S_+\rangle_{q\bar q}
  \,=\,
  -\frac{e - 1}{\sqrt{e}}\,
  T_{q\bar q}
  \,\simeq\,
  -1.04219\, T_{q\bar q}\,.
\end{equation}
We then verified this numerically in both a $q\bar q$ system and in
$qg \bar q$ systems, where in the latter we integrate over the
rapidity of the soft gluon $g$, with $p_{tg}/Q \simeq e^{-10}$, giving
\logbook{}{panscales-pc-clean/delta-scalar-pt-results/2025-06-long-v-pt/long-v-pt-consistency-check.txt}
\begin{subequations}
  \label{eq:fuzzy-cone-results}
  \begin{align}
    \langle \delta \tilde S_+\rangle_{q\bar q} \,/\, T_{q\bar q}  &= -1.04211 \pm 0.00055\,,
    \\
    \langle \delta \tilde S_+\rangle_{qg\bar q} \,/\, T_{qg\bar q} &= -1.04174 \pm 0.00082\,,
\end{align}
\end{subequations}
both in good agreement with Eq.~(\ref{eq:fuzzy-cone-expectation}).
Thus we see that the relation between average transverse and longitudinal
gluer-induced shifts is preserved in systems with a soft gluon. 
\logbook{}{See
  panscales-pc-clean/delta-scalar-pt-results/2025-06-long-v-pt/long-v-pt-consistency-check.txt
  for the raw numbers}
It is enough to have verified this for a single smooth boundary to
know that the relation will hold for arbitrary patches, because the
relation between any two patches (smooth or not) only involves an
integral over a finite range of rapidities, where only the emitted
transverse component contributes, not the longitudinal recoil along
the $q\bar q$ direction.

\subsection{Numerical evaluation of $\cR$}
\label{sec:supp:numerical-R}

Given the surprising simplicity of Eq.~(\ref{eq:Rsoln}), it is
important to verify it numerically.
To do so, we start with a $q\bar q$ system, and use a perturbative
parton shower to add a cloud of soft gluons between scales
$p_{t,\min}$ and $p_{t,\max}\ll Q$.
Summing over all the perturbative dipoles, we then probe the
effect of adding a much softer gluer at a single scale $k_{\perp n}
\ll p_{t,\min}$.
As an observable we use the transverse momentum in a central rapidity
bin with respect to the $q\bar q$ direction.
For the shower itself, we use the transverse momentum-ordered
PanGlobal shower~\cite{FerrarioRavasio:2023kyg} of the PanScales
code~\cite{vanBeekveld:2023ivn}, version~0.3, in its 
split-dipole-frame, $\beta=0$ variant, PG$_{\beta=0}^\text{sdf}$.

We have two approaches to probe the effect of the gluer: direct
insertion together with a limit as in Eq.~(\ref{eq:core-with-lim}),
and a semi-analytic approach, \S\ref{sec:supp:semi-analytic}.
With the choice of PG$_{\beta=0}^\text{sdf}$ for the perturbative
shower, multiple soft-gluon production is invariant under longitudinal
boosts along the $q\bar q$ axis, a property that helps ensure that
when the perturbative shower is used with finite $\as$ values,
subleading corrections are identical between the two gluer insertion
approaches, as we will confirm below in
\S\ref{sec:supp:results-fixed-lambda}.

\subsubsection{Semi-analytic approach}
\label{sec:supp:semi-analytic}

\begin{figure}
  \centering
  \includegraphics[width=0.5\textwidth]{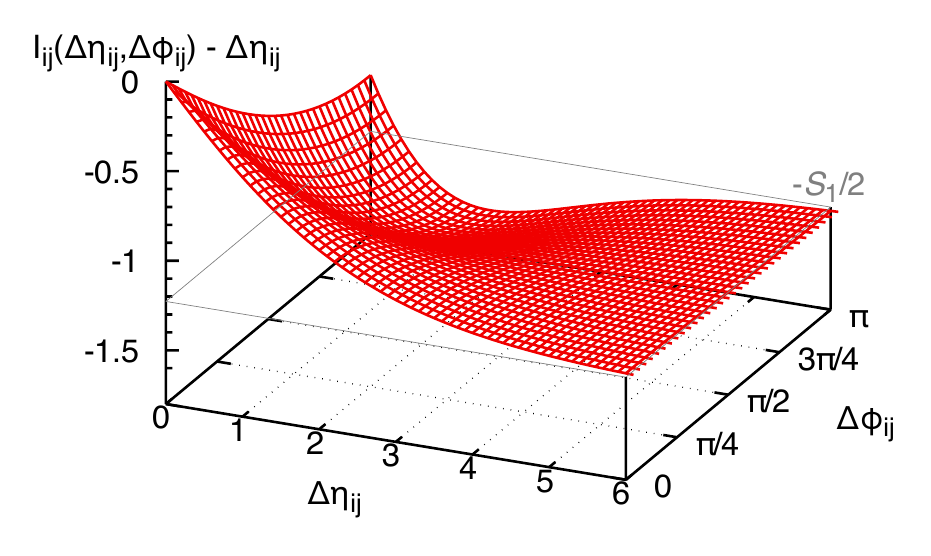}
  \caption{2d grid of $I_{ij}(\Delta\eta_{ij},\Delta\phi_{ij})$, with 
    $\Delta\eta_{ij}$ subtracted to as to better visualise the
    asymptotic behaviour, where the result (after subtraction) tends
    to $-\cS_1/2$, shown as the grey plane outline.
  }
  \label{fig:2d-grid}
\end{figure}

In the semi-analytic approach used for
Fig.~\ref{fig:all-order-many-lambdaf},
we aim to determine the effect of gluer
emission on the transverse momentum per unity rapidity, i.e.\
observables akin to $S_t(Y)$, Eq.~(\ref{eq:SY-def}).
In the main text, we showed $\rho_{qg\qbar}(\eta,\eta_g)$ for a
$qg\qbar$ system (Fig.~\ref{fig:one-emission}), but ultimately only
needed its integral over $\eta_g$, Eq.~(\ref{eq:R-integral}).
We can apply an analogous strategy for more complex system.
Specifically, we consider a generic dipole $ij$ and use an approach
similar to that of \S\ref{sec:supp:S1} to determine an analytic
function that can be integrated with adaptive Gaussian integration and
that is equivalent to
\begin{equation}
  \label{eq:Iij-def}
  I_{ij}(\Delta\eta_{ij},\Delta\phi_{ij}) =
  \int_{-\infty}^{+\infty} d\eta_n
  \rho_{ij}(\eta_n,\eta_i,\eta_j, \Delta \phi_{ij})\,,
\end{equation}
where $\rho_{ij}(\eta_n,\eta_i,\eta_j, \Delta \phi_{ij})$ is the
analogue of $\rho_{qg\qbar}(\eta,\eta_g)$ but now for the $ij$ dipole
(with the $\phi_n$ dependence already integrated).
We then tabulate $I_{ij}$ as a function of $\Delta\eta_{ij}$ and
$\Delta\phi_{ij}$, with the result shown in Fig.~\ref{fig:2d-grid}.
Note that at large $\Delta\eta_{ij}$ separations,
$I_{ij}(\Delta\eta_{ij},\Delta\phi_{ij})$ tends to $\Delta\eta_{ij} -
\tfrac12 \cS_1$, independently of $\Delta\phi_{ij}$, because the
neighbourhood of $i$ is unaffected by the specific location of $j$,
and so the analysis of \S\ref{sec:supp:S1} becomes valid.
For small $ij$ separations $I_{ij}(\Delta\eta_{ij},\Delta\phi_{ij})$
tends to zero, as is required for compatibility with the fact that a
collimated $ij$ dipole emits little soft radiation.

Because of the longitudinal boost invariance of the configurations of
the perturbative soft gluon cloud, dipoles are uniformly distributed
in rapidity along the $q\bar q$ direction.
As a result, when determining the effect of gluer emission on the
transverse momentum flow at a given rapidity, it is only the integral of
$\rho_{ij}(\eta,\eta_i,\eta_j,\Delta \phi_{ij})$ over the dipole
rapidity, $(\eta_i+\eta_j)/2$, that matters.
Accordingly, we are free to replace
$\rho_{ij}(\eta,\eta_i,\eta_j,\Delta \phi_{ij})$ with any function
that yields the same integral when integrating over
$(\eta_i+\eta_j)/2$. We take the specific form
\begin{equation}
  \label{eq:rho-replacement}
  \rho_{ij}(\eta,\eta_i,\eta_j,\Delta \phi_{ij}) \to
  \bar \rho_{ij}(\eta,\eta_i,\eta_j,\Delta \phi_{ij})
  = \Theta(\eta_i < \eta < \eta_j) + \frac12 \left[\delta(\eta-\eta_i) +
    \delta(\eta-\eta_j) \right]
  \left[I_{ij}(\Delta\eta_{ij},\Delta\phi_{ij}) - \Delta\eta_{ij} \right]
\end{equation}
For a $qi$ dipole with a quark along the positive $z$ axis we take
\begin{equation}
  \label{eq:rho-replacement-qi}
  \rho_{qi}(\eta,\eta_i,) \to
  \bar \rho_{qi}(\eta,\eta_i)
  = \Theta(\eta > \eta_i) - \frac14\cS_1 \delta(\eta-\eta_i)\,,
\end{equation}
and analogously for a $\bar q i$ along the negative $z$ axis.

To obtain the semi-analytic results in
Figs.~\ref{fig:all-order-lambdaf-04} and
\ref{fig:all-order-many-lambdaf}, we then sample over configurations
of perturbative soft-gluon clouds and then determine the following
average 
\begin{equation}
  \label{eq:avg-semi-anl}
  \langle \delta S_t(Y)\rangle
  =
  T_{q\qbar} \left \langle
    \sum_{ij \in \text{dipoles}} \int_{-Y}^Y d\eta \, \bar
    \rho_{ij}(\eta,\eta_i,\eta_j,\Delta \phi_{ij})
    \right \rangle_\text{soft-gluon configs.}
\end{equation}
over the soft-gluon clouds.

The results shown in Fig.~\ref{fig:2d-grid} are available in
computer-readable form, and through a \texttt{C++} interface, on
request from the authors.

\subsubsection{Comparing different gluer probes at fixed $\lambda$}
\label{sec:supp:results-fixed-lambda}

\begin{figure}[t!]
  \centering
  \includegraphics[width=0.5\columnwidth]{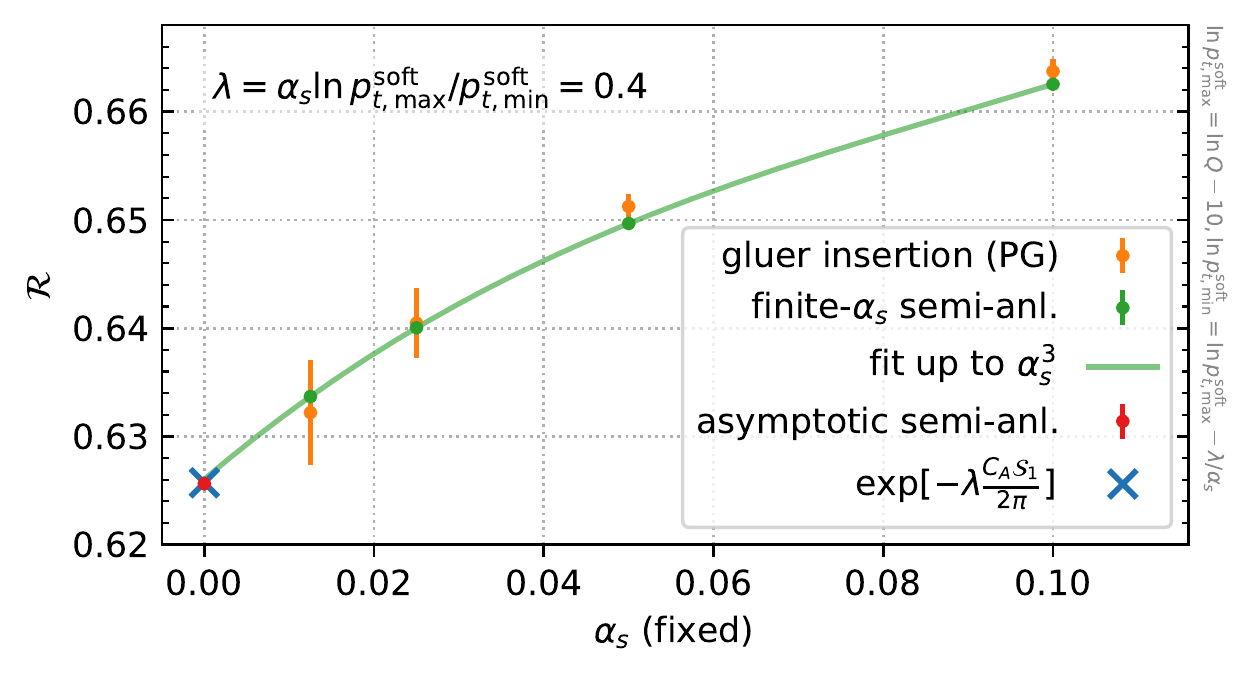}  
  \caption{
    Numerical all-order results for $\cR$ where the perturbative soft gluon cloud has been
    produced at several finite (fixed) $\as$ values and a transverse momentum
    range $\ln p_{t,\max}^\text{soft}/Q = -10$ down to $\ln p_{t,\max}^\text{soft}/Q = -10 -
    0.4/\as$, as well as at asymptotically small $\as$ in the
    semi-analytic approach.
  }
  \label{fig:all-order-lambdaf-04}
\end{figure}

Eq.~(\ref{eq:Rsoln}) is a purely single-logarithmic ($\as^n \ln^n
Q/\mu_\np$) result, whereas a shower mixes in subleading logarithms.
Here we follow the PanScales approach of examining several $\as$
values while holding fixed
$\lambda=\as \ln p_{t,\max} / p_{t,\min}=0.4$, using a fixed coupling
for simplicity, and then examining the $\as \to 0$ limit.
We choose $p_{t,\max} \ll Q$, so as to eliminate subleading effects
associated with the limited phase space for hard gluons.

Direct insertion of the gluer according to the generalisation of
Eq.~(\ref{eq:core-with-lim}) is the most flexible way of probing gluer
effects: it can be used with any observable, in both the $2$ and
$3$-jet regions and even with a massive gluer.
It is the method we have used in Figs.~\ref{fig:3-jet-anomdim-result}
and \ref{fig:mass-scheme}.
Applying it in the $2$-jet case yields the orange points of
Fig.~\ref{fig:all-order-lambdaf-04}.
While the method is flexible, the integration over the gluer
kinematics leads to significant Monte Carlo statistical errors.
Furthermore, rounding errors make it difficult to probe very small
values of $k_{\perp n}/p_{t,\max}$,  as needed for both the $\as \to
0$ and $k_{\perp n} \to 0$ limits.

In contrast, the semi-analytic approach, which exploits the fact that
the perturbative gluon cloud is uniform in rapidity at single
logarithmic accuracy, does not have these limitations.
With finite $\as$ choices, it yields the green points of
Fig.~\ref{fig:all-order-lambdaf-04}, which, for each given $\as$, are
consistent with the orange (gluer insertion) points to within
statistical uncertainties.
At the largest $\as$ values, these are well below a percent.
The semi-analytic approach can also be run at asymptotically small
$\as = 10^{-9}$ and correspondingly large
$\ln p_{t,\max} / p_{t,\min}$ (red dot).
The green points' extrapolation, the green line, is consistent with
that asymptotic run and with Eq.~(\ref{eq:Rsoln}) (blue cross).

Fig.~\ref{fig:all-order-lambdaf-04} thus validates the asymptotic
limit of the semi-analytic method that we used across many $\lambda$
values in Fig.~\ref{fig:all-order-many-lambdaf} to confirm
Eq.~(\ref{eq:Rsoln}).
Furthermore, the consistency of the two approaches at finite $\as$
values, and of the extrapolation with the asymptotic result, helps
provide confidence in the interpretation of
Figs.~\ref{fig:3-jet-anomdim-result} and \ref{fig:mass-scheme}, which
both relied on finite-$\as$ studies.

\subsection{Mass schemes}
\label{sec:supp:mass-schemes}

Calculations of non-perturbative power corrections typically use
massless partons: indeed, if one has a massive gluer as an
intermediate step, a typical full calculation will account for the
decay of that massive gluer, according to the well-known double-soft
matrix element, which leads to the Milan
factor~\cite{Dokshitzer:1997iz,Dokshitzer:1998pt,Dasgupta:1998xt,Dasgupta:1999mb},
a pure number that multiplies the result obtained with a single massless parton.
The Milan factor is universal
across linear observables.

In practice, however, experiments measure massive hadrons.
It has long been understood that different definitions of an event
shape observable, for example the choice to use the modulus of a
3-vector or an energy, will lead to $\Lambda/Q$ non-perturbative
corrections that differ~\cite{Salam:2001bd} (see also
\cite{Mateu:2012nk}).
What's more, the difference between various mass schemes appears to
come with its own anomalous dimension, which differs from that
calculated here for massless gluer emission.
A question that naturally arises is whether there is an ``ultimate''
scheme for the treatment of massive hadrons for which the relevant
anomalous dimension is just that calculated in this letter.

First, let us recall the main schemes for treating hadron masses, and
which might be good candidates for the ultimate scheme.
\begin{itemize}
\item $P$-scheme: every particle $i$ is replaced with a massless particle
  with $E_i$ set equal to $|\vec p_i|$; the centre of mass energy in
  the normalisation of the event shape is calculated from the sum of all
  the updated particle momenta.
  Ref.~\cite{Salam:2001bd} found that this scheme breaks the $c_V$ scaling of
  the non-perturbative coefficient for different event shapes as
  calculated with a massless gluer, e.g.\ the ratio of the
  $C$-parameter and thrust non-perturbative corrections is no longer
  $3\pi/2$.
  Even though the breaking is numerically small in practice (the exact breaking
  depends on the distribution of $m/k_t$ for the hadrons), that
  suggests that this scheme may be disfavoured.
  
\item $E$-scheme: every particle $i$ is replaced with a massless particle,
  with $\vec p_i$  multiplied by $E_i/|\vec p_i|$.
  In the original event's centre of mass frame, the sum of $3$-momenta
  may no longer add up to zero, and so the event is boosted back into
  the centre-of-mass frame.
  Ref.~\cite{Salam:2001bd} found that this preserves the $c_V$ scaling
  found in the massless calculation, independently of the distribution
  of $m/k_t$, and so is a good candidate scheme.

\item $D$-scheme: every massive particle is replaced by two massless
  particles, corresponding to an effective 2-body decay for all
  massive particles.
  This scheme too preserves the $c_V$ scaling found in the massless
  calculation and so appears to be as good a candidate as the
  $E$-scheme.
\end{itemize}
Ref.~\cite{Mateu:2012nk} also introduced a range of further potential
schemes.
Besides the question of the scaling of the $c_V$, there are some
general principles that strongly favour the $D$-scheme: the
conservation of momentum in the decay implies that the relation
between transverse and $+/-$ light-cone components along any given
dipole's two directions will be preserved.
As discussed in \S\ref{sec:supp:perp-v-long}, this is one of the
elements that is relevant to the all-order exponentiation of the
anomalous dimension.
Viewed another way, for asymptotically soft gluers, the insertion of a
single massive gluer that then decays gives identical kinematics for
the $q\bar q$ system as compared to the sequential insertion of two
massless gluers with the same kinematics as the gluer decay products.
Conversely, the $E$ and $P$ schemes both effectively introduce a
global element to the redistribution of longitudinal momentum with
respect to any given dipole direction, and so one might expect that
they will not reproduce various properties that we rely on for the
derivations in the massless case.

To explore this question in greater depth, we considered an explicit
kinematic map for insertion of a massive gluer.
For a dipole of mass $M$, consisting of particles with massless
momenta $\ptilde_i$ and $\ptilde_j$, and a non-perturbative gluer of
transverse momentum $k_{\perp n}$ and mass $m_n$, the map is
\begin{subequations}
  \begin{align}
    \label{eq:pl-massive}
    k_n^\mu
    &=
      z_+ \ptilde_i^\mu  + z_-  \ptilde_j^\mu  + k_{\perp n}^\mu\,,
    \\
    p_i^\mu
    &=
      (1-z_+) \ptilde_i^\mu + \frac{k_{\perp n}^2}{M^2(1-z_+)} \ptilde_j^\mu - k_{\perp n}^\mu\,,
    \\
    p_j^\mu
    &=
      \left(1 -z_- -  \frac{k_{\perp n}^2}{M^2(1-z_+)} \right)
      \ptilde_j^\mu\,,
  \end{align}
\end{subequations}
with
\begin{equation}
  \label{eq:kappa-ij-massive}
  z_+ = \sqrt{k_{\perp n}^2+m_n^2} \, \sqrt{\frac{ \tilde{s}_j}{{\tilde s}_{ij}\tilde{s}_i}}  e^{ \bar\eta},  \quad
  z_- = \sqrt{k_{\perp n}^2+m_n^2} \, \sqrt{\frac{ \tilde{s}_i}{{\tilde s}_{ij}\tilde{s}_j}}  e^{-\bar\eta} .
\end{equation}

With this map for the massive gluer insertion, we can then investigate
how all-order rescaling factor $\cR$ of Eq.~(\ref{eq:Rsoln}), when
extended to massive gluers,
depends on the ratio
of the gluer mass $m_n$ to its transverse momentum $k_{\perp n}$.
We evaluate the rescaling factor $\cR$ for the
non-perturbative gluer effect, evaluated using the same structure of
soft-gluon cloud as used in the $\as=0.1$ points of
Fig.~\ref{fig:all-order-lambdaf-04}.
Specifically, we start from a $q\bar q$ event and add a perturbative
soft-gluon cloud with transverse momenta $p_t$ in the range
$10 < \ln Q/p_t < 10 + \lambda/\as$.
We then probe how the $C$-parameter changes on inserting a massive
gluer, as compared to the integrated gluer effect in a pure $q\bar q$
event without the soft-gluon cloud.
That gives us $\cR$.

Fig.~\ref{fig:mass-scheme} of the main text shows $\cR$ for several values of
$m_n/k_{\perp n}$, in each of the three mass schemes.
Clearly for $m_n/k_{\perp n} = 0$, all schemes must agree with the
result in Fig.~\ref{fig:all-order-lambdaf-04} at the corresponding
$\as$ value, and one sees that they do.
For the $E$ and $P$ schemes, the anomalous scaling factor $\cR$
depends on $m_n/k_{\perp n}$.
Assuming this same pattern carries through to massive hadrons, this
would prevent us from predicting linear power corrections in terms of
a single $T_{q\bar q}$ parameter, since we would also need to know the
distribution of $m/k_{\perp}$ for the hadrons in order to determine the
rescaling factor $\cR$ at any given $\lambda$.
In contrast, in the $D$ scheme, the rescaling factor $\cR$ is
independent of the gluer mass, suggesting that the simple form
Eq.~(\ref{eq:Rsoln}) should hold independently of the distribution of
hadron masses, and that the $\Lambda/Q$ hadronisation corrections can
be represented in terms of a single non-perturbative parameter, i.e.\
a single effective $T_{q\bar q}$.

\begin{figure}
  \centering
  \includegraphics[width=0.5\textwidth,page=2]{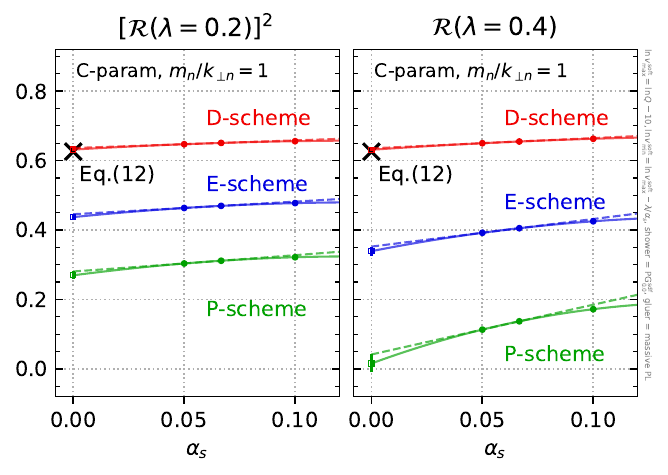}
  \caption{Calculation of $\cR$ with massive gluer
    insertion in 3 schemes for the treatment of hadron masses, shown
    as a function of different finite $\as$ values, together with the
    extrapolation for $\as \to 0$.
    Left: $[\cR(\lambda=0.2)]^2$, right
    $\cR(\lambda=0.4)$.
    The plot embodies two tests: whether the $\as \to 0$ extrapolation
    (point at $\as=0$) agrees with the
    massless calculation (black cross);
    and, if it doesn't, whether it still exponentiates, in which case
    the extrapolations  would agree with between the two
    panels.
    The solid lines show quadratic extrapolations based on all three
    $\as$ points, while the dashed lines show a linear extrapolation based
    on the two smallest $\as$ points.
    The plot thus confirms that the $D$-scheme gives results that are
    consistent with the massless calculation, while the $E$ and
    $P$-schemes do not, and furthermore do not exponentiate.\\
  }
  \label{fig:mass-schemes}
\end{figure}

Fig.~\ref{fig:mass-schemes} shows a further study.
The left and right-hand panels display $\cR$ for the
$C$-parameter for each of two values of $\lambda=\{0.2,0.4\}$, now
with a fixed $m_n/k_{\perp n}=1$, and plotted as a function of $\as$ so that
one can explore the $\as \to 0$ limit.
The right-hand panel, for $\lambda=0.4$, can be directly compared to
the massless gluer calculation of Fig.~\ref{fig:all-order-lambdaf-04}.
If the anomalous scaling exponentiates, the left-hand panel, showing
$\cR^2$ for $\lambda=0.2$, should show an effect that is identical to
the right-hand panel.
As anticipated on general grounds, and as expected already from
Fig.~\ref{fig:mass-scheme}, the $D$-scheme results are consistent with
the massless limit and, so, with exponentiation in the $\as \to 0$ limit.
In contrast, the $E$-scheme and $P$-scheme clearly do not satisfy
exponentiation and they show significantly stronger anomalous scaling
than the $D$-scheme results, as was already seen in
Fig.~\ref{fig:mass-scheme}.

Our focus here has been on identifying whether there is a mass scheme
whose anomalous dimension coincides with the massless result.
One comment is that while our anomalous dimension in the $D$-scheme is
negative (the power correction decreases faster than $1/Q$), the
anomalous dimensions for differences between schemes has been found to
be positive Refs.~\cite{Salam:2001bd,Mateu:2012nk} (see Fig.~5 and
Fig.~8 respectively), i.e.\ the difference between mass schemes
decreases more slowly than $1/Q$.
Given that $P-E$ and $E-D$ differences are negative, and that the
anomalous dimension for the difference is larger than for the $D$
scheme, this implies that in the $E$ and $P$ schemes the hadronisation
will change sign at sufficiently large $Q$.
This was actually observed long ago in Fig.~8 of \cite{Salam:2001bd}.
The $\as\to0$ limit of Fig.~\ref{fig:mass-schemes} also suggests
proximity to a sign-change in the $P$ scheme for $\lambda=0.4$.
In future work, it would clearly be of interest to further examine the
differences between mass schemes within our approach here, and more
explicitly compare the results with those of
Refs.~\cite{Salam:2001bd,Mateu:2012nk}.

Taking a wider perspective, we should emphasise that there are
multiple potential considerations in choosing a mass scheme.
Here we have focused on the perspective of how one might
estimate hadronisation corrections.
Depending on the scope and purpose of any given measurement, other
considerations might be more relevant.
One notable criterion might be how closely a given scheme reflects
what is actually measured in a given experimental setup, e.g.\ the
relative use of tracks versus calorimetry for charged and neutral
particles, the availability particle-identification, and the
differences in calorimetric response between baryons and anti-baryons.
%

\subsection{Continuous bin edges for energy correlators}
\label{sec:supp:cont-bin}

In our studies of gluer insertion for the EEC with a condition
$|\cos\theta|<1/2$, we found that at large 
$Q$ values and very small $k_{\perp n}$, there were large
statistical uncertainties.
These appear to be associated with situations where a perturbative
gluon (with transverse momentum $p_t$) is close to the $\theta$
boundary, within an angle of order $k_{\perp n}/p_t$.
A gluer with a large $z_+$ light-cone momentum fraction can go to the
other side of the boundary, changing the observable by an amount that
can be almost as large as $p_t/Q \gg k_{\perp n}/Q$.
Ultimately such effects average out to be small: sometimes the
perturbative gluon is outside the boundary and sometimes it is inside,
changing the sign of the effect, and the effect is only present in the
rare situations where the perturbative gluon is close to the boundary.
Nevertheless the effect leaves a large trace in the dispersion of the
result.

To work around this issue, we found it useful to adopt a continuous
bin, in analogy with Eq.~(\ref{eq:fuzzy-cone}) 
\begin{equation}
  \label{eq:cont-binC}
  \text{EEC} (|\cos\theta|\lesssim 1/2) = \sum_{i,j} \frac{E_iE_j}{Q^2} f(|\cos\theta|)
  \qquad
    f(c) = \Theta\left(0.4 < c <
    0.6\right)\times\left(\frac{0.6-c}{0.2}\right)  + \Theta\left(c < 0.4 \right)\,.
\end{equation}
This greatly reduced the statistical fluctuations.
It is our understanding that such continuous bins are sometimes used
also to improve the stability of fixed-order perturbative
calculations, though this point is not necessarily always explicitly
discussed.

\subsection{Study of Q dependence of shifts in distributions}
\label{sec:supp:more-distributions}

\begin{figure}[tb]
  \centering
  \includegraphics[width=0.45\columnwidth,page=1]{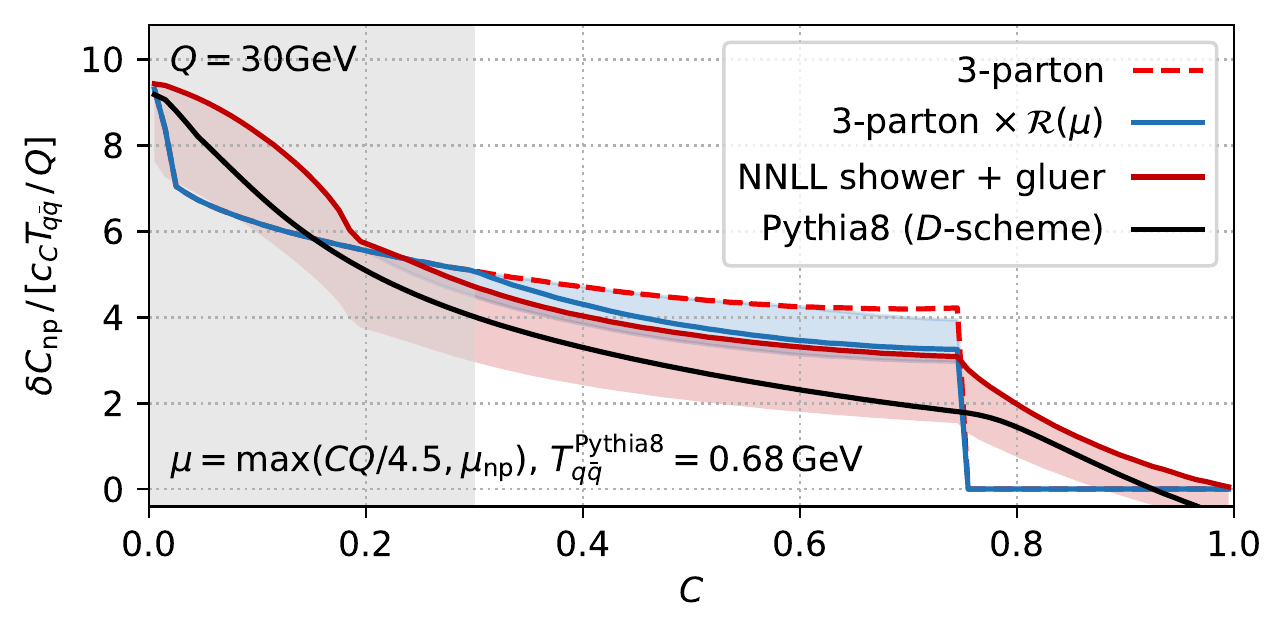}
  \includegraphics[width=0.45\columnwidth,page=7]{plotting/anomdim_v_fullshower_ktcut2.pdf}  
  \includegraphics[width=0.45\columnwidth,page=2]{plotting/anomdim_v_fullshower_ktcut2.pdf}
  \includegraphics[width=0.45\columnwidth,page=8]{plotting/anomdim_v_fullshower_ktcut2.pdf}  
  \includegraphics[width=0.45\columnwidth,page=3]{plotting/anomdim_v_fullshower_ktcut2.pdf}
  \includegraphics[width=0.45\columnwidth,page=9]{plotting/anomdim_v_fullshower_ktcut2.pdf}  
  \includegraphics[width=0.45\columnwidth,page=4]{plotting/anomdim_v_fullshower_ktcut2.pdf}
  \includegraphics[width=0.45\columnwidth,page=10]{plotting/anomdim_v_fullshower_ktcut2.pdf}    
  \caption{
    Non-perturbative shifts in the $C$-parameter (left) and thrust
    (right) for multiple values of the centre-of-mass energy $Q$.
    The set of curves is the same as in
    Fig.~\ref{fig:pythia-comparison}.
    Specifically, the dark red line corresponds to gluer insertion
    into a shower that runs down to a scale $\mu_\np = 2\GeV$.
    That insertion is carried out with the PL shower and a scale
    $k_{\perp n}\simeq 0.0015\GeV$, while the red band shows the
    uncertainty when the insertion switches to $k_{\perp n}\simeq
    0.7\GeV$ with the PG shower.
    For the dark blue line, based on Eq.~(\ref{eq:Rsoln}), the band
    shows the impact of varying $\mu$
    by a factor of $1.5$ up ($2$ down) around the central choice as
    shown in the figure.
    The motivation for the central scale choice is shown discussed in
    the text.
    As was the case in Fig.~\ref{fig:pythia-comparison}, the grey area indicates
    $CQ/4.5 < \mu_\np = 2\GeV$, where we freeze $\mu$ to
    $\mu_{\np}$ and where the perturbative shower cutoff in the gluer insertion
    curve starts to have an effect.
  }
  \label{fig:differentQ}
\end{figure}

We determine hadronisation corrections in Pythia8 as follows.
Firstly, we run Pythia8 including hadronisation and determine the
hadron-level value of an observable from the hadrons.
We then inspect the event record to identify the last partons before
the Lund hadronisation stage, using the Pythia8's internal
option \texttt{isFinalPartonLevel}.
We use those partons to calculate a parton-level value for the
observable.
We then take all events in a given bin of parton-level value for the
observable and determine the average of the difference between the
hadron and parton-level values of the observable.
That average is the shift that enters a given bin for our Pythia8
results.

Fig.~\ref{fig:differentQ} shows versions of
Fig.~\ref{fig:pythia-comparison} for a wide range of $Q$ values,
$Q=\{30, 91.2, 1000, 10000\}\GeV$ and for both the $C$-parameter
(left) and the thrust (right).
Starting with the $C$-parameter, one sees that as $Q$ is increased
above $91.2\GeV$, there is quite remarkable agreement between three
different sets of the results: the analytic $\cR(\mu)$ rescaling, the
direct gluer insertion into the shower, and Pythia.
Instead, at the lowest $Q$ value of $30\GeV$, the uncertainties
associated with the insertion scale and map for the gluer (red band)
grow significant.
Those uncertainties are important in terms of obtaining consistency
with Pythia's hadronisation correction.

For the thrust, two main remarks are in order: the uncertainty
associated with the insertion scale and map for the gluer (red band)
is more substantial than for the $C$-parameter, for reasons that we
have yet to elucidate.
Furthermore the difference between the rescaled 3-parton result (dark
blue) and the gluer insertion (dark red) remains significant even at
very high $Q$ values.
This difference is a potential measure of further higher-order
effects.
Given that the leading-order thrust distribution goes to zero linearly
at $\tau=\frac13$, the fractional NLO correction to the distribution there
diverges as $1/(\tau-\frac13)$, and one might therefore expect the thrust to be less
perturbatively stable than the $C$-parameter, where such a power divergence does
not occur.
However we have not been able to translate this observation into a
more quantitative expectation for the difference between the two
approaches.
Nevertheless, overall, Fig.~\ref{fig:differentQ} suggests that the
thrust is a less stable observable than the $C$-parameter when it
comes to estimating non-perturbative corrections, a consideration that
should probably be taken into account in strong-coupling determinations.

A final comment concerns the choice of $\mu$ in the $\cR(\mu)$
rescaling of the 3-parton result.
For a single soft gluon at zero rapidity, the transverse momentum of the
gluon is given by $p_t=CQ/3$ or $p_t=\tau/Q$.
As a default choice for $\mu$ we take this $p_t$ divided by $1.5$, on
the grounds that the soft resummation will start somewhere below the
actual hardest gluon.
We choose an uncertainty band such that $\mu$ does not exceed this
$p_t$.

